\documentclass[12pt,letterpaper]{article}

\usepackage[left=1in,right=1in,top=1in,bottom=1in]{geometry}

\usepackage{setspace}

\usepackage{amsmath,amssymb,amsthm,mathtools,bm}

\usepackage{graphicx}
\usepackage{booktabs}
\usepackage{threeparttable}
\usepackage{longtable}
\usepackage{array}
\usepackage{multirow}
\usepackage{caption}
\usepackage{subcaption}
\usepackage{float}
\usepackage{placeins}
\usepackage{makecell}

\usepackage{enumitem}
\usepackage{xcolor}
\usepackage{appendix}
\usepackage{pdflscape}
\usepackage{csquotes}

\usepackage{chngcntr}
\counterwithin{figure}{section}
\counterwithin{table}{section}

\usepackage[authoryear]{natbib}

\usepackage[colorlinks=true,linkcolor=blue,citecolor=blue,urlcolor=blue]{hyperref}
\usepackage[nameinlink,capitalize]{cleveref}

\theoremstyle{definition}

\title{The P behind Q:\\Empirical Evidence from\\Physical Drift in Put--Call Parity}
\author{Useong Shin\thanks{
		Sogang Business School, Sogang University (Seoul, Korea).\\
		ORCID: \href{https://orcid.org/0009-0003-0197-9003}{0009-0003-0197-9003}\\
		Email: \texttt{useong@sogang.ac.kr}
}}
\date{\today}

\begin{document}
	
	\maketitle
	\thispagestyle{empty}
	
	\begin{flushleft}
		\textbf{\small JEL:} G12; G13; G14\\
		\textbf{\small Keywords:} carry gap; put--call parity; physical drift; path risk; P--Q tension; limits to arbitrage
	\end{flushleft}
	
	\noindent\textbf{Acknowledgments:}
	I am grateful to Michele Azzone (Politecnico di Milano) for generously sharing OIS data, for guidance on implementing the implied-discount-factor pipeline, and for detailed feedback on earlier drafts. All remaining errors are my own.
	
	\begin{abstract}
		Put--call parity is a terminal-payoff identity, but its enforcement is capital-using.
		I study the carry gap, the annualized wedge between option-implied and OIS discount factors, in SPX and RUT index options.
		Quoted parity is tightly compressed, while the synthetic--traded forward channel leaves a systematic wedge.
		I interpret this wedge as an implementation premium under finite arbitrage capital.
		A drift-preserving GBM term, \(r\hat{\mu}\tau\), improves in-sample and leave-one-year-out fit, especially in SPX.
		The evidence suggests that physical drift enters not option payoffs, but the process enforcing risk-neutral parity.
	\end{abstract}
	
	\pagenumbering{arabic}
	
	\newpage
	
	\section{Introduction}
	\label{sec:intro}
	
	Put--call parity is a terminal-payoff identity. For European calls and puts with the same strike and maturity, the call--put spread is pinned down by the corresponding forward position and discount factor \citep{Stoll69,BS73,Merton73,GG74}. The physical-measure expected return of the underlying has no direct role in this risk-neutral pricing argument. Put--call parity is therefore closed by a \(Q\)-measure relation, not by the \(P\)-measure drift of the underlying.
	
	This paper studies the implementation layer behind that \(Q\)-measure relation. In a frictionless benchmark, quoted put--call parity, the traded-futures channel, and the option-implied discount factor would all be jointly aligned with the benchmark money-market curve. With finite arbitrage capital, however, these equalities need not be jointly costless to enforce. The \(P\) behind \(Q\), in this paper, is the physical price path along which capital must be committed to maintain a risk-neutral parity relation.
	
\begin{figure}[H]
		\centering
		\includegraphics[width=6.5in]{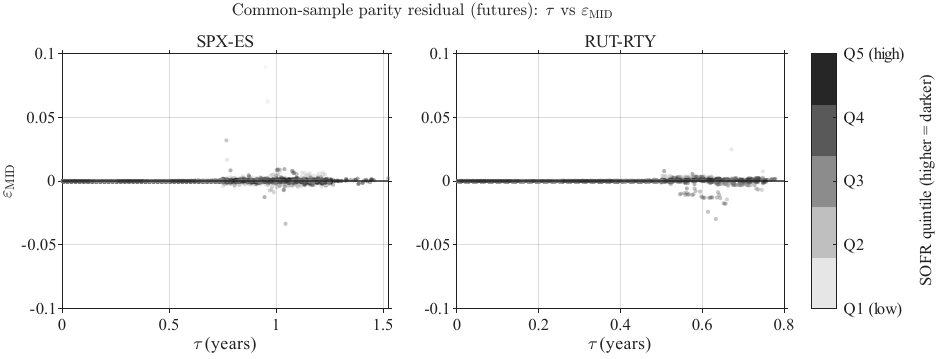}
		\includegraphics[width=6.5in]{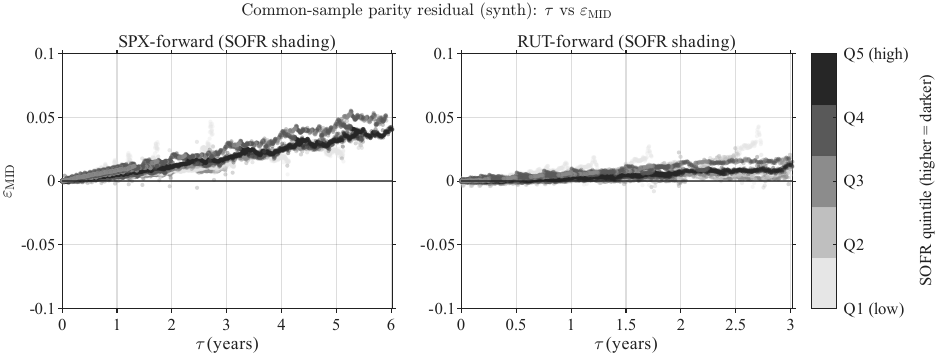}
		\caption{Price-space residuals under two forward identification channels. The figure is reproduced from \citet*{Shin26}. The upper panels use traded futures as the forward input; residuals are tightly compressed around zero. The lower panels recover the synthetic forward from the option cross-section while imposing the same OIS discount factor; a systematic synthetic--traded forward wedge remains.}
		\label{fig:cpresidsynth}
	\end{figure}

	As documented in \citet*{Shin26}, this distinction is visible in SPX and RUT index options. Quoted put--call parity residuals against traded futures are tightly compressed in price space, but the OIS-discounted synthetic-forward channel is not equally aligned with the traded-futures channel. Figure~\ref{fig:cpresidsynth}, reproduced from \citet*{Shin26}, illustrates this contrast: the traded-futures channel leaves residuals close to zero, whereas the option-cross-sectional synthetic-forward channel leaves a systematic wedge relative to traded futures when the same OIS discount factor is imposed.
	
	The central empirical fact is therefore not that put--call parity visibly fails. It is that quoted parity can be nearly closed while a related OIS-discounted synthetic--traded forward wedge remains open. I interpret this wedge as an implementation wedge: the residual cost of maintaining compressed parity when arbitrage capital, funding, margin capacity, and execution liquidity are finite. This interpretation does not require a literal violation of spot--futures parity; the conservative claim is that the two forward-identification channels are not perfectly aligned after implementation costs are taken into account.
	
	I measure this wedge using the carry gap, the annualized difference between the option-implied discount factor and the benchmark OIS discount factor. The object is measured in carry space because small price-space residuals need not imply costless enforcement. \citet*{Shin26} shows that the SPX and RUT carry gap is related to an \(r\sigma\sqrt{\tau}\) GBM support-capital term, trading frictions, and broad financial conditions. The \(r\sigma\sqrt{\tau}\) term summarizes the path-dependent capital burden implied by zero-drift Brownian motion.
	
	\FloatBarrier
	\FloatBarrier
	
	This paper asks whether the same implementation-risk logic changes once physical drift is preserved. If the carry gap reflects the cost of maintaining parity under finite arbitrage capital, it should depend not only on volatility and funding rates, but also on the direction of the physical price path. The reason is that parity can be enforced from two sides. The \(+C-P-F\) and \(-C+P+F\) positions enforce the same terminal relation, but their futures-leg exposures have opposite signs before maturity. Physical drift therefore creates asymmetric variation-margin and capital-support burdens across enforcement directions.
	
	\FloatBarrier
	\begin{figure}[H]
		\centering
		\includegraphics[width=6.5in]{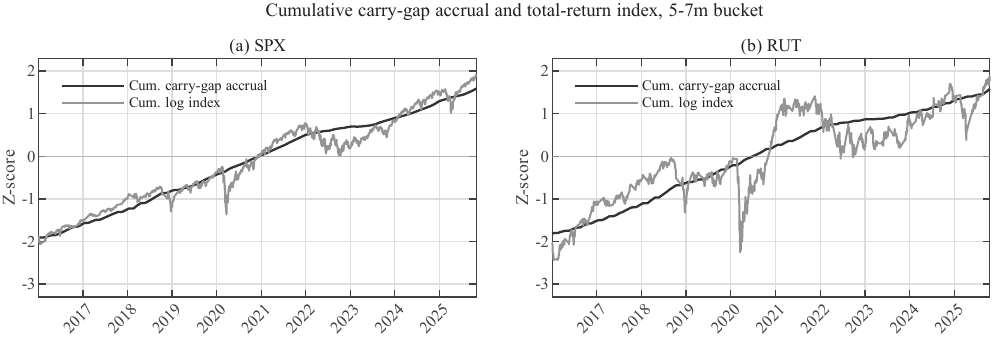}
		\caption{Cumulative carry-gap accrual and total-return index in the 5--7 month maturity bucket. For each market, the carry gap is aggregated to a daily pooled series within the maturity bucket and cumulated over time. Both series are standardized as z-scores.}
		\label{fig:cg_traj}
	\end{figure}
	\FloatBarrier
	
	Figure~\ref{fig:cg_traj} provides the motivating diagnostic. The figure is
	deliberately suggestive rather than dispositive. Indeed, the first concern is
	that the apparent comovement may be spurious: two cumulative financial series
	can look aligned because of common trends, regime shifts, or persistent state
	variables. I therefore use the figure only to motivate the test, not as
	evidence for the mechanism. The empirical question is whether a prior
	physical-drift proxy explains the carry gap after the diffusion-based
	implementation-risk block is controlled for.
	
	The drift-preserving extension follows from the same support-capital argument. In the zero-drift case, the implementation burden scales as \(r\sigma\sqrt{\tau}\). Preserving drift adds, to first order, a directional margin-burden component proportional to \(r\mu\tau\). Because the true physical drift \(\mu\) is unobserved, I use \(\hat{\mu}\), a prior rolling-OLS slope of the log total-return index, and implement the drift component as \(r\hat{\mu}\tau\). Throughout the paper, \(\hat{\mu}\) is a reduced-form historical-drift proxy, not an observed expected return.
	
	The evidence supports the drift-preserving extension, with stronger results in SPX than in RUT. Using a common 504-trading-day drift proxy, the extended model improves both in-sample fit and leave-one-year-out out-of-sample performance. In SPX, pooled OOS \(R^2\) rises from 0.221 to 0.310. In RUT, it rises from 0.171 to 0.188. The gain is not concentrated at a single optimized lookback window; it appears over a stable medium-horizon plateau. The drift loading is positive and statistically significant in both markets under HAC(21) inference, but the economic pass-through is larger in SPX.
	
	These results do not imply that physical drift prices option payoffs directly. Put--call parity remains a risk-neutral terminal-payoff identity. The claim is narrower: physical price direction enters the margin and capital burden of the process that enforces risk-neutral parity.
	
	The paper makes three contributions. First, it reframes the carry gap as an implementation wedge that remains visible in carry space even when quoted put--call parity is tightly compressed in price space. Second, it derives and tests a drift-preserving extension of the GBM implementation-risk term: \(r\sigma\sqrt{\tau}\) captures diffusion-driven path risk, while \(r\hat{\mu}\tau\) captures directional drift burden. Third, it shows that this drift term has incremental explanatory power for SPX and RUT index options after controlling for diffusion path risk, trading frictions, and financial conditions.
	
	The paper proceeds as follows. \Cref{sec:lit} reviews the related literature. \Cref{sec:theory} presents the GBM path-risk term and the drift-preserving extension. \Cref{sec:data} describes the data and empirical specifications. \Cref{sec:base} summarizes the baseline regression. \Cref{sec:ext} reports the drift-extended results. \Cref{sec:discussion} and \Cref{sec:conclusion} discuss the economic interpretation and conclude.
	
	\section{Related Literature}
	\label{sec:lit}
	
	This paper connects to three strands of literature: put--call parity and
	option-implied discounting, limits to arbitrage and implementation costs, and
	the relation between risk-neutral objects and physical-measure information. The
	common distinction is simple. Put--call parity is exact as a terminal-payoff
	identity, but the trading strategy that enforces it before maturity is
	path-dependent and capital-using. I study this distinction in carry space by
	asking whether the capital burden of parity enforcement is related to a
	physical-drift proxy.
	
	Put--call parity is a foundational no-arbitrage relation for European options
	\citep{Stoll69,BS73,Merton73,GG74}. Empirical work shows that observed
	deviations should not be read mechanically as failures of no-arbitrage logic:
	transaction costs, execution frictions, short-sale constraints, and market
	microstructure can all affect measured parity residuals
	\citep{KR79,AT01}. I retain this implementation view, but shift the object of
	analysis. Rather than asking whether the call--put price difference violates the
	terminal payoff identity, I study the carry gap: the annualized wedge between
	the option-implied discount factor and the benchmark OIS discount factor. The
	question is therefore not whether put--call parity holds as a terminal relation,
	but whether enforcing that relation before maturity leaves a systematic,
	state-dependent cost.
	
	This object is closest in measurement to the option-implied discounting
	literature. \citet{BG86} infer implied interest rates from option prices, while
	\citet{AB21} estimate option-implied discount factors from European put--call
	parity and compare them with OIS benchmarks. Closely related work by
	\citet*{Shin26} applies this approach to SPX and RUT options and documents that
	the resulting carry gap is related to a GBM path-risk term, trading frictions,
	and broad financial conditions. The present paper extends that structure by
	preserving physical drift in the same support-capital logic. The zero-drift term
	\(r\sigma\sqrt{\tau}\) captures diffusion-driven path risk; the drift-preserving
	extension adds a directional component proportional to \(r\mu\tau\), implemented
	empirically as \(r\hat{\mu}\tau\).
	
	The economic interpretation builds on the limits-to-arbitrage literature.
	Since \citet{SV97}, this literature has emphasized that arbitrage is performed
	by finite-capital intermediaries, not unconstrained representative investors.
	Funding constraints, margin requirements, mark-to-market losses, and liquidation
	risk can make trades costly to carry even when terminal payoffs appear locked
	\citep{GV02,BP09,MP12}. In options markets, \citet{ORW04} show that arbitrage
	constraints can be linked to put--call parity deviations. My setting is
	deliberately different. I study European-style index options, where visible
	price-space parity residuals against traded futures are already tightly
	compressed. The question is not why an obvious arbitrage violation remains
	open, but why a carry-space wedge remains after the visible residual has largely
	been closed.
	
	This view is also close to evidence on implementation costs in other
	no-arbitrage relations. In covered interest parity, for example, deviations can
	persist when balance-sheet capacity, funding conditions, and intermediary
	constraints make enforcement costly \citep{DTV18}. The market and measurement
	object are different, but the economic logic is similar. A no-arbitrage relation
	may remain valid at the terminal-payoff level while its enforcement is affected
	by capital constraints, margin requirements, funding conditions, and interim
	price paths.
	
	Finally, the paper relates to work connecting risk-neutral objects with
	physical-measure information. Existing studies show that option-implied
	\(Q\)-measure objects can contain information about physical probabilities,
	variance risk premia, or expected returns \citep{BP09,Ross15,Martin17}. My
	question is complementary. I do not recover physical probabilities or expected
	returns from option prices, and I do not argue that physical drift directly
	prices option payoffs. Put--call parity remains a risk-neutral terminal-payoff
	relation. Instead, I ask whether the carry-space residual associated with that
	relation is exposed to physical-measure price dynamics through the arbitrageur's
	path-dependent margin and capital burden.
	
	In sum, the contribution is to test a drift-preserving implementation-risk
	channel inside a clean parity setting. Where prior work studies price-space
	deviations, implied discount rates, or zero-drift implementation risk, I derive
	and test an \(r\hat{\mu}\tau\) extension of the GBM support-capital term. The
	evidence should be read not as a failure of risk-neutral parity, but as
	reduced-form evidence that the capital-using mechanism enforcing a
	risk-neutral relation can be exposed to physical-measure price dynamics.
	
	\section{Path-Risk Term and Drift-Extended Specification}
	\label{sec:theory}
	
	This section develops the GBM-based empirical specification for the carry gap.
	The central idea is simple: put--call parity is exact at maturity, but the
	trade that enforces it must survive before maturity. The enforcing position is
	exposed to daily settlement, variation margin, interim losses, funding costs,
	and finite-capital constraints. The GBM terms are therefore not auxiliary
	controls. They are reduced-form summaries of the capital burden created by
	carrying a parity-enforcement trade through the pre-maturity path.
	
	The derivation starts from a zero-drift Brownian approximation to interim
	profit and loss. In that case, support capital is driven by diffusion risk and
	scales with \(\sigma\sqrt{\tau}\). Multiplying this support-capital scale by an
	opportunity-cost rate gives the \(r\sigma\sqrt{\tau}\) path-risk term.
	Preserving physical drift in the same calculation leaves the diffusion term in
	place, but adds a first-order directional burden proportional to \(\mu\tau\).
	The two components therefore take the reduced-form forms
	\(r\sigma\sqrt{\tau}\) and \(r\mu\tau\).
	
	The theoretical \(\mu\) and the empirical \(\hat{\mu}\) should be kept separate.
	In the derivation, \(\mu\) denotes the true physical drift or expected return.
	In the data, this object is unobserved. I therefore use \(\hat{\mu}\), a prior
	rolling-OLS slope of past log total-return paths. The drift term is not meant
	to estimate the true expected return structurally. It is a reduced-form test of
	whether an observable historical-drift proxy aligns with the directional margin
	burden of parity enforcement.
	
	For notational simplicity, \(t\) indexes observations in the
	market--date--maturity panel. Daily variables such as \(OIS_t\), \(NFCI_t\),
	\(Vol_{i,t}\), and \(\hat{\mu}_{i,t}\) are matched to all maturity cells
	observed on the same date. The variable \(\tau_{i,t}\) denotes time to maturity
	in years.
	
	\subsection{A Motivating Example}
	\label{subsec:bob_example}
	
	Consider Bob, an arbitrageur who observes a put--call parity wedge and enters
	\[
	+C-P-F.
	\]
	On paper, the trade looks almost self-contained. The call and put form a
	synthetic forward, and the futures leg offsets much of the directional
	exposure. In a highly idealized margin calculation, the net required margin can
	look small relative to the gross legs. Bob appears to have opened an almost
	costless parity-enforcement trade.
	
	The problem is that maturity is not the whole story. The options mainly pay at
	expiration, but the futures leg settles every day. If the index rises and Bob's
	short futures leg loses 1,000 dollars, he must post cash immediately. He may
	borrow it, draw down internal liquidity, or use cash that could have been
	deployed elsewhere. In all cases, the cash has a funding cost, an opportunity
	cost, or both. If the index falls the next day and Bob receives 2,000 dollars
	through variation margin, he can redeploy the surplus in a money-market
	account. The sign of the cash flow has changed, but the implementation problem
	has not disappeared. Until maturity, Bob must finance cash shortfalls and
	redeploy temporary surpluses.
	
	Now consider Alice, who enters the market on a day when the parity wedge has
	the opposite sign. She enforces the same terminal relation from the other side:
	\[
	-C+P+F.
	\]
	Her terminal payoff is also pinned down by put--call parity, but her path risk
	is the mirror image of Bob's. If the index rises, Alice's long futures leg
	receives variation margin while Bob's short futures leg pays it. If the index
	falls, the cash-flow burden reverses. Thus Bob and Alice enforce the same
	terminal relation, but they do not face the same pre-maturity cash-flow path.
	
	This is where physical drift matters. A persistent positive drift does not
	change the terminal put--call parity identity, but it makes the short-futures
	enforcement direction more likely to face repeated margin calls. A persistent
	negative drift shifts the burden to the long-futures enforcement direction.
	Drift therefore acts asymmetrically across the two sides of parity enforcement.
	It determines which side is more likely to consume funding capacity before the
	deterministic terminal payoff is reached.
	
	This is the distinction the model tries to capture. The parity payoff may be
	deterministic at maturity, but the path to maturity is not capital-free. Daily
	settlement turns a terminal no-arbitrage identity into a pre-maturity
	cash-management problem. The GBM path-risk term summarizes the support capital
	required to survive random path variation. The drift extension summarizes the
	directional imbalance in that burden across the two enforcement directions.
	
	The approximation below is not a literal model of exchange margin rules or of
	every institutional detail of futures settlement. It is a first-order
	support-capital approximation. Under zero-drift Brownian motion, the relevant
	object is the running maximum of adverse interim P\&L, which scales with
	\(\sigma\sqrt{\tau}\). Multiplying by an opportunity-cost rate \(r\) gives the
	diffusion component \(r\sigma\sqrt{\tau}\). If the futures leg has nonzero
	physical drift, the same support-capital logic adds a first-order directional
	component proportional to \(\mu\tau\), up to a constant and a sign determined by
	the enforcement direction. In the empirical specification, the unobserved
	\(\mu\) is replaced by the prior rolling-OLS proxy \(\hat{\mu}\).
	
	\subsection{Baseline Diffusion Path-Risk Term}
	\label{subsec:diffusion_path_risk}
	
	The functional form of the GBM term follows from the support-capital problem
	illustrated above. A synthetic long forward combined with a short futures
	contract pays
	\[
	(S_T-K)+(F_0-S_T)=F_0-K
	\]
	at maturity, so the terminal payoff is deterministic. Before maturity,
	however, the futures leg is settled through the margin account. An adverse
	price path can require capital injections before the deterministic terminal
	payoff is reached.
	
	Consider the normalized interim profit-and-loss process
	\[
	X_u=\sigma B_u,
	\qquad u\in[0,\tau],
	\]
	where \(B_u\) is standard Brownian motion and \(\sigma\) is annualized
	volatility. Let \(\ell_u\) denote the minimum support capital required to keep
	the position solvent:
	\[
	X_u+\ell_u\ge 0
	\qquad \text{for all } u\in[0,\tau],
	\qquad \ell_0=0.
	\]
	The smallest nondecreasing process satisfying this constraint is
	\[
	\ell_u=\sup_{0\le s\le u}(-X_s)^+ .
	\]
	
	By a standard property of Brownian motion,
	\[
	\mathbb{E}[\ell_u]
	=
	\sigma\sqrt{\frac{2u}{\pi}} .
	\]
	Thus required support capital is proportional to volatility and increases with
	the square root of time. The average capital burden over the life of the trade is
	\[
	\bar{B}(\tau)
	=
	\frac{1}{\tau}\int_0^\tau \mathbb{E}[\ell_u]\,du
	=
	\frac{2}{3}\sigma\sqrt{\frac{2\tau}{\pi}} .
	\]
	If the opportunity cost of tied-up capital is summarized by a rate-like object
	\(r_t\), the representative path-risk scale is
	\[
	r_t\bar{B}(\tau)
	=
	r_t
	\frac{2}{3}\sigma\sqrt{\frac{2\tau}{\pi}} .
	\]
	
	The baseline GBM term is this \(r\sigma\sqrt{\tau}\) diffusion burden translated
	into basis points. For market \(i\in\{\mathrm{SPX},\mathrm{RUT}\}\),
	\begin{equation}
		\label{eq:gbm_diffusion_term_definition_kr}
		\begin{aligned}
			GBM_{i,t}^{\sigma,OIS,xY}
			&=
			10^4
			\cdot
			\frac{OISxY_t}{100}
			\cdot
			\frac{2}{3}
			\cdot
			\frac{Vol_{i,t}}{100}
			\cdot
			\sqrt{\frac{2\tau_{i,t}}{\pi}},
			\qquad x\in\{1,10\},
			\\[0.5em]
			Vol_{i,t}
			&=
			\begin{cases}
				VIX_t, & i=\mathrm{SPX},\\
				RVX_t, & i=\mathrm{RUT}.
			\end{cases}
		\end{aligned}
	\end{equation}
	The \(x=1\) component captures short- to medium-horizon funding conditions,
	whereas the \(x=10\) component captures longer-horizon opportunity cost. The
	term is not a structural estimate. It is a reduced-form support-capital scale
	proportional to \(r\sigma\sqrt{\tau}\).
	
	\subsection{Drift-Preserving Path-Risk Extension}
	\label{subsec:drift_preserving_extension}
	
	I next preserve physical drift in the support-capital calculation. The terminal
	parity payoff remains deterministic, but the pre-maturity margin path now
	depends on the direction of the futures exposure. Drift therefore adds a
	directional burden to the diffusion-based path-risk term.
	
	Let the normalized interim profit-and-loss process be
	\[
	X_u^{(q)}
	=
	q\mu u+\sigma B_u,
	\qquad u\in[0,\tau],
	\]
	where \(\mu\) is the true physical drift or expected return and
	\(q\in\{+1,-1\}\) denotes the direction of the parity-enforcement position.
	The trades \(+C-P-F\) and \(-C+P+F\) enforce the same terminal parity relation
	from opposite sides, but their futures-leg exposures, and therefore their drift
	exposures, have opposite signs.
	
	The minimum support capital is the cumulative capital needed to offset negative
	interim P\&L:
	\[
	\ell_u^{(q)}
	=
	\sup_{0\le s\le u}
	\left(-X_s^{(q)}\right)
	=
	\sup_{0\le s\le u}
	\left(-q\mu s-\sigma B_s\right).
	\]
	Because the expression includes \(s=0\), it is nonnegative. Since \(-B_s\) is
	again standard Brownian motion, \(\ell_u^{(q)}\) has the same distribution as
	the running maximum of Brownian motion with drift parameter \(-q\mu\).
	
	Define the auxiliary process
	\[
	Y_s^{(m)}
	=
	m s+\sigma \widetilde{B}_s,
	\qquad s\in[0,u],
	\]
	where \(\widetilde{B}_s\) is standard Brownian motion and \(m\) is an auxiliary
	drift parameter. Let
	\[
	M_u^{(m)}
	=
	\sup_{0\le s\le u}Y_s^{(m)} .
	\]
	Then
	\[
	\ell_u^{(q)}
	\overset{d}{=}
	M_u^{(-q\mu)} .
	\]
	
	A first-order expansion of the expected running maximum around \(m=0\) gives
	\[
	\mathbb{E}\!\left[M_u^{(m)}\right]
	=
	\sigma\sqrt{\frac{2u}{\pi}}
	+
	\frac{m u}{2}
	+
	O\!\left(
	\frac{m^2u^{3/2}}{\sigma}
	\right).
	\]
	Substituting \(m=-q\mu\),
	\[
	\mathbb{E}\!\left[\ell_u^{(q)}\right]
	=
	\sigma\sqrt{\frac{2u}{\pi}}
	-
	\frac{q\mu u}{2}
	+
	O\!\left(
	\frac{\mu^2u^{3/2}}{\sigma}
	\right).
	\]
	
	Averaging over the life of the trade,
	\[
	\bar{B}^{(q)}(\tau)
	=
	\frac{1}{\tau}
	\int_0^\tau
	\mathbb{E}\!\left[\ell_u^{(q)}\right]du ,
	\]
	so
	\[
	\bar{B}^{(q)}(\tau)
	=
	\frac{2}{3}\sigma\sqrt{\frac{2\tau}{\pi}}
	-
	\frac{q\mu\tau}{4}
	+
	O\!\left(
	\frac{\mu^2\tau^{3/2}}{\sigma}
	\right).
	\]
	Multiplying by the opportunity cost \(r_t\) gives
	\[
	r_t\bar{B}^{(q)}(\tau)
	=
	r_t
	\frac{2}{3}\sigma\sqrt{\frac{2\tau}{\pi}}
	-
	\frac{q}{4}r_t\mu\tau
	+
	O\!\left(
	\frac{r_t\mu^2\tau^{3/2}}{\sigma}
	\right).
	\]
	
	This decomposition shows that the zero-drift GBM term is the diffusion-driven
	support-capital burden,
	\[
	r\sigma\sqrt{\tau},
	\]
	while preserving drift adds a first-order directional margin-burden component,
	\[
	r\mu\tau .
	\]
	The constant \(-q/4\) and the sign of the enforcement direction can be absorbed
	into the regression coefficient, because the observed carry gap reflects the
	net imbalance of two-sided parity-enforcement pressure. I therefore use the
	drift burden as a reduced-form regressor.
	
	The true \(\mu\) is unobserved. I replace it empirically with \(\hat{\mu}\), a
	prior rolling-OLS slope of past log total-return paths, and implement the
	theoretical drift burden as \(r\hat{\mu}\tau\). This substitution does not
	identify the true expected return. It only supplies an observable
	historical-drift proxy for the directional margin-burden term. In the main
	specification, I scale the drift burden only by the one-year OIS rate:
	\begin{equation}
		\label{eq:gbm_drift_term_definition_kr}
		GBM_{i,t}^{\hat{\mu},OIS,1Y}
		=
		10^4
		\cdot
		\frac{OIS1Y_t}{100}
		\cdot
		\hat{\mu}_{i,t}^{ann}
		\cdot
		\tau_{i,t}.
	\end{equation}
	I do not include a 10-year-OIS-scaled drift term in the main extended
	specification. The goal is not to search over additional long-rate-scaled drift
	components, but to test the most parsimonious empirical implication of the
	theory: whether a directional \(r\hat{\mu}\tau\) margin-burden proxy adds
	explanatory power on top of the OIS-based diffusion path-risk block.
	
	Specifically, I estimate the rolling OLS slope of the past \(n\)-trading-day log
	total-return path and annualize it as
	\[
	\hat{\mu}_{i,t}^{ann}
	=
	252\hat{\mu}_{i,t}^{daily}.
	\]
	To avoid look-ahead bias, the drift proxy used at date \(t\) is computed only
	from information available through \(t-1\). Thus
	\(GBM_{i,t}^{\hat{\mu},OIS,1Y}\) is not an observed expected return, but a
	reduced-form implementation-risk regressor based on a prior physical-drift
	proxy extrapolated from past medium-horizon price paths. Constants, scaling
	conventions, and the enforcement-side sign are absorbed by the regression
	coefficient.
	
	The derivation therefore motivates two empirical blocks: a diffusion burden
	scaling with \(r\sigma\sqrt{\tau}\), and a drift burden scaling with
	\(r\mu\tau\). Since \(\mu\) is unobserved, the empirical specification uses
	\(r\hat{\mu}\tau\).
	
	\subsection{Drift Asymmetry and Carry-Gap--Index Comovement}
	\label{subsec:drift_asymmetry}
	
	The two-sided enforcement logic helps interpret the comovement between
	cumulative carry-gap accrual and the total-return index in
	Figure~\ref{fig:cg_traj}. A positive physical drift makes the short-futures
	enforcement direction more likely to consume funding capacity, whereas a
	negative drift shifts the burden toward the long-futures direction. The
	terminal parity relation is unchanged, but the pre-maturity margin burden is
	not symmetric across enforcement directions.
	
	This mechanism does not structurally identify the comovement in
	Figure~\ref{fig:cg_traj}. The figure is only suggestive, and cumulative
	financial series can comove for spurious reasons. The empirical test is
	therefore narrower: whether \(r\hat{\mu}\tau\), with \(\hat{\mu}\) measured
	only from prior return paths, explains the carry gap after the diffusion
	path-risk block is controlled for.
	
	\subsection{Baseline and Drift-Extended Specifications}
	\label{subsec:drift_regression_specs}
	
	I now define the empirical regressions. Because SPX and RUT differ in liquidity,
	investor composition, and microstructure, the main specification is estimated
	separately by market. This avoids imposing a common slope vector across the two
	markets and allows the same path-risk structure to load with different
	intensities.
	
	The baseline specification is the OIS-based GBM regression. For each market
	\(i\in\{\mathrm{SPX},\mathrm{RUT}\}\),
	\begin{equation}
		\label{eq:baseline_gbm_spec_kr}
		\begin{aligned}
			CG_{i,t}^{bp}
			=
			{}&
			\alpha_i
			+
			\phi_{1,i}GBM_{i,t}^{\sigma,OIS,1Y}
			+
			\phi_{10,i}GBM_{i,t}^{\sigma,OIS,10Y}
			\\
			&+
			\beta_i
			\frac{BA_{i,t}^{\mathrm{med}}}{\tau_{i,t}}
			+
			\gamma_i NFCI_t
			+
			\varepsilon_{i,t}.
		\end{aligned}
	\end{equation}
	Here \(GBM^{\sigma,OIS,1Y}\) and \(GBM^{\sigma,OIS,10Y}\) are diffusion
	path-risk terms scaled by the one-year and ten-year OIS rates,
	\(BA_{i,t}^{\mathrm{med}}/\tau_{i,t}\) captures trading frictions, and
	\(NFCI_t\) proxies for broad financial conditions.
	
	The drift-extended specification adds a single OIS-one-year-scaled drift term,
	\(GBM_{i,t}^{\hat{\mu},OIS,1Y}\):
	\begin{equation}
		\label{eq:market_specific_gbm_drift_spec_kr}
		\begin{aligned}
			CG_{i,t}^{bp}
			=
			{}&
			\alpha_i
			+
			\phi_{1,i}GBM_{i,t}^{\sigma,OIS,1Y}
			+
			\phi_{10,i}GBM_{i,t}^{\sigma,OIS,10Y}
			\\
			&+
			\psi_iGBM_{i,t}^{\hat{\mu},OIS,1Y}
			+
			\beta_i
			\frac{BA_{i,t}^{\mathrm{med}}}{\tau_{i,t}}
			+
			\gamma_i NFCI_t
			+
			\varepsilon_{i,t}.
		\end{aligned}
	\end{equation}
	The only difference between the baseline and drift-extended specifications is
	the inclusion of \(GBM_{i,t}^{\hat{\mu},OIS,1Y}\). The coefficient \(\psi_i\)
	therefore measures whether the physical-drift proxy aligns with the carry gap
	after controlling for diffusion path risk, trading frictions, and broad
	financial conditions.
	
	\subsection{Economic Interpretation of the Regressors}
	\label{subsec:regressor_interpretation}
	
	The specification decomposes the implementation burden of parity enforcement
	into diffusion path risk and directional drift burden. The diffusion term,
	\[
	GBM^{\sigma}\sim r\sigma\sqrt{\tau},
	\]
	captures support capital induced by random path variation. The drift term,
	\[
	GBM^{\hat{\mu}}\sim r\hat{\mu}\tau,
	\]
	proxies for the directional margin burden associated with prior physical price
	trends. The true drift is \(\mu\); the empirical regressor uses
	\(\hat{\mu}\), a historical proxy.
	
	The coefficients are reduced-form pass-through loadings, not structural
	primitives. The carry gap reflects the net result of two-sided
	parity-enforcement pressure, so constants, signs, and partial offsetting across
	arbitrage directions are absorbed by the regression coefficients.
	
	The remaining controls capture implementation conditions outside the GBM block.
	
	\(BA_{i,t}^{\mathrm{med}}/\tau_{i,t}\) summarizes annualized trading frictions,
	and \(NFCI_t\) proxies for broad funding and balance-sheet conditions. The
	coefficient on \(GBM^{\hat{\mu},OIS,1Y}\) should therefore be read as the
	incremental pass-through from an observable drift proxy to the carry gap, not as
	a structural loading on the true physical drift.
	
	\section{Data and Methodology}
	\label{sec:data}
	
	This section describes the data, carry-gap construction, and empirical design.
	The analysis compares a restricted OIS-based baseline with a drift-extended
	specification. The baseline uses diffusion path-risk terms, trading frictions,
	and broad financial conditions. The extended specification adds the
	\(r\hat{\mu}\tau\) drift-burden proxy motivated in \Cref{sec:theory}.
	
	\subsection{Data and Carry-Gap Construction}
	\label{subsec:data_sample}
	
	I study European-style index options on SPX and RUT. The option data are
	minute-level NBBO quotes from ThetaData, and the sample is restricted to dates
	for which both option quotes and OIS data are available. Because these index
	options are European-style contracts, early-exercise premia do not enter the
	put--call-parity-based identification of discount factors.
	
	I construct the carry gap using the synthetic-forward identification procedure
	of \citet*{AB21}. For European calls and puts with the same strike \(K\) and
	maturity \(T\), put--call parity can be written as
	\begin{equation}
		C_t(K,T)-P_t(K,T)
		=
		B_t(T)\left(F_t(T)-K\right),
	\end{equation}
	where \(B_t(T)\) is the option-implied discount factor and \(F_t(T)\) is the
	forward value at the same maturity. Define the synthetic forward as
	\begin{equation}
		\mathcal{G}_t(K,T)=C_t(K,T)-P_t(K,T).
	\end{equation}
	Under no arbitrage, the recovered forward should be independent of strike. I
	therefore use the cross-sectional relation between synthetic forwards and
	strikes within each date--maturity cell to jointly estimate \(\hat{B}_t(T)\)
	and \(\hat{F}_t(T)\).
	
	The benchmark discount factor is constructed from the OIS curve. Let
	\(\tau_t(T)=T-t\) denote time to maturity. The carry gap is
	\begin{equation}
		CG_t(T)
		=
		\frac{1}{\tau_t(T)}
		\log\left(
		\frac{D_t^{OIS}(T)}{\hat{B}_t(T)}
		\right),
		\label{eq:carry_gap_def_kr}
	\end{equation}
	where \(D_t^{OIS}(T)\) is the OIS discount factor and \(\hat{B}_t(T)\) is the
	option-implied discount factor. I report the basis-point version,
	\begin{equation}
		CG_t^{bp}(T)=10^4\cdot CG_t(T).
		\label{eq:carry_gap_bp_def_kr}
	\end{equation}
	
	The preprocessing keeps only matched call--put pairs with the same strike and
	maturity. I remove observations with abnormally low prices, excessive
	bid--ask spreads, too few valid strikes within a date--maturity cell, or failed
	OIS curve construction. The final panel consists of market--date--maturity
	observations for which both \(\hat{B}_t(T)\) and \(D_t^{OIS}(T)\) can be
	constructed reliably.
	
	\subsection{Baseline and Drift-Extended Specifications}
	\label{subsec:baseline_method}
	
	The baseline specification is the OIS-based GBM regression. For market
	\(i\in\{\mathrm{SPX},\mathrm{RUT}\}\),
	\begin{equation}
		\label{eq:baseline_spec_data_kr}
		\begin{aligned}
			CG_{i,t}^{bp}
			=
			{}&
			\alpha_i
			+
			\phi_{1,i}GBM_{i,t}^{\sigma,OIS,1Y}
			+
			\phi_{10,i}GBM_{i,t}^{\sigma,OIS,10Y}
			\\
			&+
			\beta_i
			\frac{BA_{i,t}^{\mathrm{med}}}{\tau_{i,t}}
			+
			\gamma_i NFCI_t
			+
			\varepsilon_{i,t}.
		\end{aligned}
	\end{equation}
	Here \(GBM^{\sigma,OIS,1Y}\) and \(GBM^{\sigma,OIS,10Y}\) are the
	\(r\sigma\sqrt{\tau}\) diffusion path-risk terms scaled by one-year and
	ten-year OIS rates. The variable \(BA_{i,t}^{\mathrm{med}}/\tau_{i,t}\)
	summarizes annualized trading frictions, and \(NFCI_t\) proxies for broad
	financial conditions.
	
	The extended specification adds the drift-preserving term from
	\Cref{sec:theory}. The theoretical drift \(\mu\) is unobserved. I therefore use
	\(\hat{\mu}\), a prior rolling-OLS slope of past log total-return paths, as an
	empirical proxy. To avoid look-ahead bias, the proxy used at date \(t\) is
	computed only from information available through \(t-1\). Let
	\(\hat{\mu}^{daily}_{i,t}\) denote this rolling daily slope. The annualized
	proxy is
	\begin{equation}
		\hat{\mu}^{ann}_{i,t}
		=
		252\hat{\mu}^{daily}_{i,t}.
	\end{equation}
	The proxy is not an observed expected return. It is a historical drift measure
	used to test the directional margin-burden channel.
	
	In the main specification, the drift burden is scaled only by the one-year OIS
	rate:
	\begin{equation}
		\label{eq:drift_term_method_kr}
		GBM_{i,t}^{\hat{\mu},OIS,1Y}
		=
		10^4
		\cdot
		\frac{OIS1Y_t}{100}
		\cdot
		\hat{\mu}_{i,t}^{ann}
		\cdot
		\tau_{i,t}.
	\end{equation}
	I do not include a ten-year-OIS-scaled drift term. The goal is to test the
	most parsimonious implication of the derivation: whether an
	\(r\hat{\mu}\tau\) margin-burden proxy adds explanatory power beyond the
	\(r\sigma\sqrt{\tau}\) path-risk block.
	
	The drift-extended specification is
	\begin{equation}
		\label{eq:extended_spec_data_kr}
		\begin{aligned}
			CG_{i,t}^{bp}
			=
			{}&
			\alpha_i
			+
			\phi_{1,i}GBM_{i,t}^{\sigma,OIS,1Y}
			+
			\phi_{10,i}GBM_{i,t}^{\sigma,OIS,10Y}
			\\
			&+
			\psi_iGBM_{i,t}^{\hat{\mu},OIS,1Y}
			+
			\beta_i
			\frac{BA_{i,t}^{\mathrm{med}}}{\tau_{i,t}}
			+
			\gamma_i NFCI_t
			+
			\varepsilon_{i,t}.
		\end{aligned}
	\end{equation}
	The coefficient \(\psi_i\) measures whether the historical-drift proxy aligns
	with the carry gap after controlling for diffusion path risk, trading
	frictions, and broad financial conditions.
	
	\subsection{Drift Horizon and Evaluation Method}
	\label{subsec:mu_horizon_method}
	
	The rolling drift proxy depends on the lookback horizon \(n\). I estimate the
	drift-extended specification over multiple values of \(n\) and examine how
	in-sample and leave-one-year-out out-of-sample performance vary across
	horizons. The purpose is not to rely on a single pointwise optimum. It is to
	check whether the drift proxy has stable explanatory power over a broad horizon
	band.
	
	The main specification uses \(n=504\) trading days, roughly two trading years.
	This horizon is naturally interpreted as a medium-horizon physical-drift proxy
	rather than a short-term return-noise measure. I use the same horizon for SPX
	and RUT to restrict specification flexibility and keep the empirical design
	common across markets.
	
	I evaluate the baseline and drift-extended specifications along three
	dimensions. First, I report in-sample fit using \(R^2\), adjusted \(R^2\), RMSE,
	and MAE. Second, I conduct leave-one-year-out out-of-sample tests and report
	OOS \(R^2\), RMSE, MAE, and fitted--actual correlation. Third, I report
	date-based HAC(Newey--West) coefficient inference with a maximum lag of 21
	trading days.
	
	The central test is whether the \(r\hat{\mu}\tau\) term explains the level and
	time variation of the carry gap after controlling for \(r\sigma\sqrt{\tau}\)
	path risk, trading frictions, and financial conditions. The term should be
	read as a reduced-form implementation-risk component, not as a generic return
	predictor or as a structural estimate of the true drift \(\mu\).
	
	\section{Baseline GBM Regression}
	\label{sec:base}
	
	Before introducing the drift-preserving term, I summarize the restricted GBM
	benchmark. The purpose is to establish the diffusion-based path-risk block
	against which the drift extension is evaluated. The baseline corresponds to
	equation~\eqref{eq:baseline_spec_data_kr} and uses OIS-scaled diffusion GBM
	terms, trading frictions, and broad financial conditions. The next section
	adds only one regressor, \(GBM^{\hat{\mu},OIS,1Y}\), to this same structure.
	
	\subsection{In-Sample Benchmark}
	
	\FloatBarrier
	\begin{table}[H]
		\centering
		\onehalfspacing
		\footnotesize
		\caption{In-sample fit of the baseline GBM regression}
		\label{tab:base_is_summary_kr}
		\resizebox{6.5in}{!}{
			\begin{tabular}{lcccccc}
				\toprule
				Specification & Obs. & Trading days & \(R^2\) & Adj.\ \(R^2\) & RMSE (bp) & MAE (bp) \\
				\midrule
				Pooled common + SPX dummy & 48{,}013 & 2{,}456 & 0.309 & 0.309 & 13.57 & 9.26 \\
				SPX separate              & 29{,}368 & 2{,}456 & 0.312 & 0.312 & 13.20 & 8.68 \\
				RUT separate              & 18{,}645 & 2{,}455 & 0.281 & 0.281 & 13.95 & 10.10 \\
				\bottomrule
			\end{tabular}
		}
	\end{table}
	\FloatBarrier
	
		Table~\ref{tab:base_is_summary_kr} reports the in-sample fit. The pooled
	common-slope specification has an \(R^2\) of 0.309, while the market-specific
	regressions produce \(R^2\) values of 0.312 for SPX and 0.281 for RUT. Separate
	slopes do not materially change the fit, suggesting that the diffusion-based
	GBM structure is broadly shared across the two markets. This is the benchmark
	that the drift term must improve upon.
	
	\subsection{Coefficient Structure}
	
	\FloatBarrier
	\begin{table}[H]
		\centering
		\onehalfspacing
		\footnotesize
		\caption{Coefficient estimates of the baseline GBM regression: HAC(21) inference}
		\label{tab:base_coef_hac21_kr}
		\begin{tabular}{lccc}
			\toprule
			Regressor 
			& Pooled common + SPX dummy 
			& SPX separate 
			& RUT separate \\
			\midrule
			
			Intercept
			& 24.901*** & 23.134*** & 24.577*** \\
			& (5.816)   & (5.713)   & (5.407)   \\
			
			\(D^{\mathrm{SPX}}\)
			& -0.985    & ---       & ---       \\
			& (0.713)   &           &           \\
			
			\(GBM^{\sigma,OIS,1Y}\)
			& -0.557*** & -0.548*** & -0.555*** \\
			& (0.148)   & (0.170)   & (0.124)   \\
			
			\(GBM^{\sigma,OIS,10Y}\)
			& 0.469***  & 0.411**   & 0.541***  \\
			& (0.151)   & (0.172)   & (0.130)   \\
			
			\(BA^{\mathrm{med}}/\tau\)
			& 0.158***  & 0.256***  & 0.130***  \\
			& (0.029)   & (0.064)   & (0.022)   \\
			
			\(NFCI\)
			& -24.598** & -25.839** & -23.961** \\
			& (10.283)  & (10.359)  & (10.013)  \\
			
			\midrule
			Trading days 
			& 2{,}456 & 2{,}456 & 2{,}455 \\
			\bottomrule
		\end{tabular}
		
		\vspace{0.5em}
		\begin{minipage}{0.95\linewidth}
			\footnotesize
			\textit{Notes:} Standard errors in parentheses are date-based HAC
			(Newey--West) standard errors with maximum lag 21 trading days.
			In the pooled common specification, \(D^{\mathrm{SPX}}\) is an indicator equal to one for SPX
			and zero for RUT, so the intercept corresponds to the RUT level.
			***, **, * denote significance at the 1\%, 5\%, and 10\% levels, respectively.
		\end{minipage}
	\end{table}
	\FloatBarrier
	
		Table~\ref{tab:base_coef_hac21_kr} shows stable coefficient signs across
	markets. \(GBM^{\sigma,OIS,1Y}\) is negative in all specifications, whereas
	\(GBM^{\sigma,OIS,10Y}\) is positive in all specifications; both remain
	statistically significant under HAC(21) inference. The opposite signs suggest
	that the GBM block is not merely a discount-rate level proxy. Instead,
	shorter- and longer-horizon opportunity-cost components load differently on
	the net carry-gap imbalance.
	
	The trading-friction variable, \(BA^{\mathrm{med}}/\tau\), is positive and
	significant throughout, consistent with wider execution costs increasing the
	carry gap. \(NFCI\) is negative and significant in all specifications,
	indicating that broad financial conditions remain related to carry-gap
	variation after the OIS-based path-risk block is controlled for.
	
	\subsection{LOYO Benchmark}
	
	\FloatBarrier
	\begin{table}[H]
		\centering
		\onehalfspacing
		\footnotesize
		\caption{LOYO out-of-sample performance of the baseline GBM regression}
		\label{tab:base_oos_summary_kr}
		\resizebox{6.5in}{!}{
			\begin{tabular}{llcccccc}
				\toprule
				Specification & Market & Mean \(R^2\) & Median \(R^2\) & Pooled \(R^2\) & Years with \(R^2>0\) & Mean corr. & Mean RMSE (bp) \\
				\midrule
				Common-market & SPX & 0.049 & 0.187 & 0.212 & 9/10 & 0.189 & 13.93 \\
				Common-market & RUT & 0.065 & 0.063 & 0.173 & 6/10 & 0.252 & 15.16 \\
				Separate      & SPX & 0.059 & 0.130 & 0.221 & 7/10 & 0.205 & 13.95 \\
				Separate      & RUT & 0.075 & 0.108 & 0.171 & 6/10 & 0.243 & 15.07 \\
				\bottomrule
			\end{tabular}
		}
	\end{table}
	\FloatBarrier
	
		Table~\ref{tab:base_oos_summary_kr} reports leave-one-year-out performance.
	The market-specific baseline has modest mean OOS \(R^2\): 0.059 for SPX and
	0.075 for RUT. This is not a high-frequency forecasting model. Even so, pooled
	OOS \(R^2\) remains positive, at 0.221 for SPX and 0.171 for RUT. The baseline
	therefore retains out-of-sample information under year-level sample splits,
	although its calibration weakens in abrupt regime transitions, especially
	around the 2020 holdout.
	
	\subsection{Coefficient-Sign Stability}
	
	\FloatBarrier
	\begin{table}[H]
		\centering
		\onehalfspacing
		\footnotesize
		\caption{LOYO coefficient-sign stability of the baseline GBM regression: HAC(21) inference}
		\label{tab:base_oos_sign_hac21_kr}
		\begin{tabular}{lcccc}
			\toprule
			Regressor
			& SPX sign
			& SPX HAC significance
			& RUT sign
			& RUT HAC significance \\
			\midrule
			Intercept
			& \(+\)\;10/10 & 10/10 at 1\%
			& \(+\)\;10/10 & 10/10 at 1\% \\
			
			\(GBM^{\sigma,OIS,1Y}\)
			& \(-\)\;10/10 & 9/10 at 5\%
			& \(-\)\;10/10 & 10/10 at 1\% \\
			
			\(GBM^{\sigma,OIS,10Y}\)
			& \(+\)\;10/10 & 8/10 at 5\%
			& \(+\)\;10/10 & 10/10 at 1\% \\
			
			\(BA^{\mathrm{med}}/\tau\)
			& \(+\)\;10/10 & 10/10 at 1\%
			& \(+\)\;10/10 & 10/10 at 1\% \\
			
			\(NFCI\)
			& \(-\)\;10/10 & 8/10 at 5\%, 10/10 at 10\%
			& \(-\)\;10/10 & 9/10 at 5\%, 10/10 at 10\% \\
			\bottomrule
		\end{tabular}
		
		\vspace{0.5em}
		\begin{minipage}{0.92\linewidth}
			\footnotesize
			\textit{Notes:} Each entry is computed from the ten leave-one-year-out
			training-sample regressions for the market-specific baseline specification.
			Standard errors are date-based HAC (Newey--West) standard errors
			with maximum lag 21 trading days.
		\end{minipage}
	\end{table}
	\FloatBarrier
	
	Table~\ref{tab:base_oos_sign_hac21_kr} reports coefficient-sign stability
	across LOYO training samples. The four non-intercept regressors have fully
	stable signs in both markets: \(GBM^{\sigma,OIS,1Y}\) is always negative,
	\(GBM^{\sigma,OIS,10Y}\) is always positive, \(BA^{\mathrm{med}}/\tau\) is
	always positive, and \(NFCI\) is always negative. The pattern also survives
	HAC(21) inference. In RUT, both GBM terms and the bid--ask term are significant
	at the 1\% level in all folds. In SPX, the one-year GBM term is significant at
	the 5\% level in 9 of 10 folds, and the ten-year GBM term in 8 of 10 folds.
	
	The HAC lag choice is conservative. The LOYO training samples contain roughly
	2,200 trading dates, for which common automatic Newey--West bandwidth rules
	imply lags below 10 trading days. The reported HAC(21) inference instead uses
	an approximately one-month autocorrelation window, so the sign stability is not
	driven by aggressive standard-error adjustment.
	
	\subsection{Summary}
	
	The restricted GBM benchmark explains a meaningful share of in-sample
	carry-gap variation, retains positive pooled OOS \(R^2\) under LOYO validation,
	and has stable coefficient signs across re-estimation folds. The next section
	therefore asks whether adding \(GBM^{\hat{\mu},OIS,1Y}\) improves an already
	informative \(r\sigma\sqrt{\tau}\)-based benchmark.
	
	\section{Empirical Results for the Drift-Extended Specification}
	\label{sec:ext}
	
	This section tests the drift-preserving extension of the baseline GBM
	specification. The extended model adds a single term,
	\(GBM^{\hat{\mu},OIS,1Y}\), to the restricted baseline. The comparison is
	therefore deliberately narrow: I ask whether the reduced-form
	\(r\hat{\mu}\tau\) component explains carry-gap variation not already captured by
	the diffusion-based \(r\sigma\sqrt{\tau}\) path-risk block, trading frictions,
	and financial conditions. I do not include an OIS-10Y-scaled drift term, because
	the goal is to test the most parsimonious empirical implication of the
	drift-preserving support-capital argument rather than to search over additional
	rate-scaled drift components.
	
	\subsection{Drift Proxy and Horizon Scan}
	\label{subsec:mu_horizon_scan}
	
	The drift-extended specification depends on the lookback horizon \(n\) used to
	construct the empirical drift proxy. In the derivation, \(\mu\) denotes the true
	physical drift or conditional expected return. In the data, this object is
	unobserved. I therefore use \(\hat{\mu}_{i,t}^{(n)}\), a prior rolling-OLS slope
	of the log total-return path. The proxy is not a structural estimate of
	\(\mu\); it is an observable historical-drift proxy used to construct the
	reduced-form \(r\hat{\mu}\tau\) term.
	
	Let \(TR_{i,t}\) denote the total-return index for market
	\(i\in\{\mathrm{SPX},\mathrm{RUT}\}\). To avoid look-ahead bias, the \(n\)-day
	proxy used at date \(t\) is computed only from information available through
	\(t-1\). Specifically, I estimate
	\begin{equation}
		\log TR_{i,t-n+\ell}
		=
		a_{i,t}^{(n)}
		+
		b_{i,t}^{(n)}\ell
		+
		u_{i,t,\ell}^{(n)},
		\qquad
		\ell=0,1,\ldots,n-1,
	\end{equation}
	and annualize the slope as
	\begin{equation}
		\hat{\mu}_{i,t}^{(n)}
		=
		252\,\hat{b}_{i,t}^{(n)}.
	\end{equation}
	Thus, \(\hat{\mu}_{i,t}^{(n)}\) is an annualized drift proxy in decimal units,
	constructed entirely from prior returns.
	
	For each horizon \(n\), I define the OIS-1Y-scaled drift term as
	\begin{equation}
		GBM_{i,t}^{\hat{\mu},OIS,1Y}(n)
		=
		10^4
		\cdot
		\frac{OIS_t^{1Y}}{100}
		\cdot
		\hat{\mu}_{i,t}^{(n)}
		\cdot
		\tau_{i,t}.
		\label{eq:mu_hat_gbm_term}
	\end{equation}
	The OIS rate is divided by 100 because it is quoted in percentage points,
	whereas \(\hat{\mu}_{i,t}^{(n)}\) enters in annualized decimal units.
	
	I estimate separate market regressions over alternative values of \(n\) and
	track both in-sample \(R^2\) and leave-one-year-out pooled out-of-sample
	\(R^2\). The horizon-scan regression adds only the drift term in
	equation~\eqref{eq:mu_hat_gbm_term} to the baseline. Because the main empirical
	specification is market-specific, the scan is conducted separately for SPX and
	RUT.
	
	\FloatBarrier
	\begin{figure}[H]
		\centering
		\includegraphics[width=6.5in]{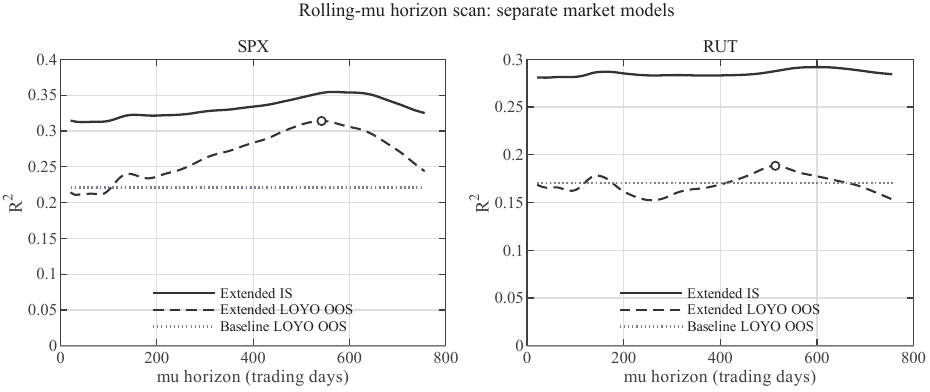}
		\caption{Rolling-\(\hat{\mu}\) horizon scan for the market-specific
			drift-extended regressions. The drift-extended specification adds only the
			OIS-1Y-scaled drift term, \(GBM^{\hat{\mu},OIS,1Y}\), to the baseline.
			Solid lines report in-sample \(R^2\), dashed lines report LOYO pooled
			out-of-sample \(R^2\), and horizontal dashed lines report the corresponding
			baseline LOYO pooled out-of-sample \(R^2\).}
		\label{fig:scan}
	\end{figure}
	\FloatBarrier
	
	Figure~\ref{fig:scan} shows that the drift-extended specification outperforms
	the baseline over a broad medium-horizon region. The pointwise OOS optima are
	close to two trading years: 542 trading days for SPX, 514 trading days for RUT,
	and 525 trading days for the two-market average. I use \(n=504\), roughly two
	trading years, as the main horizon. This choice imposes a common horizon across
	markets and is naturally interpreted as a medium-horizon drift proxy.
	
	The 504-day horizon is not a mechanical ex post optimum. Its out-of-sample cost
	is negligible: the OOS \(R^2\) is 0.310 instead of the SPX optimum of 0.314,
	0.188 instead of the RUT optimum of 0.189, and 0.249 instead of the two-market
	average optimum of 0.250. The selected horizon also lies inside the near-optimal
	plateau in all three cases. It is therefore an interpretable, common, and stable
	medium-horizon choice rather than a pointwise optimum.
	
\begin{figure}[H]
		\centering
		\includegraphics[width=6.5in]{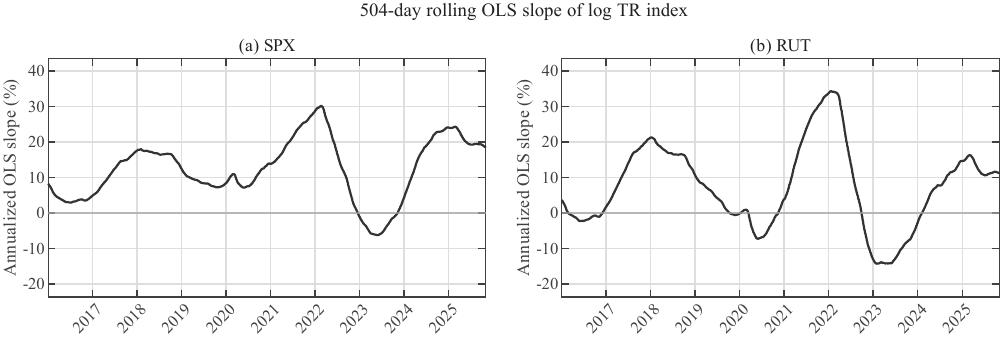}
		\caption{Drift proxy estimated from the 504-trading-day prior-only rolling
			OLS slope of the log total-return index. The slope is annualized and reported
			in percent. Panel (a) reports SPX and Panel (b) reports RUT. The sample runs
			from January 2016 to October 2025. Each slope estimate uses only information
			available before the corresponding date.}
		\label{fig:rolling}
	\end{figure}

	Figure~\ref{fig:rolling} plots the resulting 504-day prior-only rolling OLS
	slope, annualized and expressed in percent. In both SPX and RUT,
	\(\hat{\mu}_{i,t}^{(504)}\) behaves as a smooth low-frequency series rather than
	as a high-frequency return measure.
	
	\FloatBarrier
	\FloatBarrier
	
	Figures~\ref{fig:scan} and~\ref{fig:rolling} clarify the interpretation of the
	drift proxy. The explanatory gain is not concentrated in a narrow optimized
	window, and the selected 504-day series captures medium-term direction rather
	than high-frequency return noise. This supports interpreting \(r\hat{\mu}\tau\)
	as a directional margin-burden proxy in parity enforcement, not as a mechanical
	same-day return predictor.
	
	\subsection{In-Sample Results}
	\label{subsec:is_results}
	
	I next compare the in-sample performance of the drift-extended specification
	with the baseline. The extended model uses the common 504-trading-day
	\(\hat{\mu}\) proxy selected in Section~\ref{subsec:mu_horizon_scan} and adds
	only one OIS-1Y-scaled drift term to the baseline:
	\begin{align}
		CG^{bp}_{i,t}
		=
		\alpha_i
		&+
		\phi_{1,i}GBM^{\sigma,OIS,1Y}_{i,t}
		+
		\phi_{10,i}GBM^{\sigma,OIS,10Y}_{i,t}
		+
		\psi_i GBM^{\hat{\mu},OIS,1Y}_{i,t}  \notag \\
		&+
		\beta_i\frac{BA^{med}_{i,t}}{\tau_{i,t}}
		+
		\gamma_i NFCI_t
		+
		\varepsilon_{i,t}.
		\label{eq:drift_ext_final}
	\end{align}
	The added regressor is
	\begin{equation}
		GBM^{\hat{\mu},OIS,1Y}_{i,t}
		=
		10^4
		\cdot
		\frac{OIS^{1Y}_t}{100}
		\cdot
		\hat{\mu}^{ann}_{i,t}
		\cdot
		\tau_{i,t}.
		\label{eq:mu_term_final}
	\end{equation}
	Thus, the baseline and extended specifications differ only by the inclusion of
	\(GBM^{\hat{\mu},OIS,1Y}_{i,t}\).
	
	\FloatBarrier
	\begin{figure}[H]
		\centering
		\includegraphics[width=6.5in]{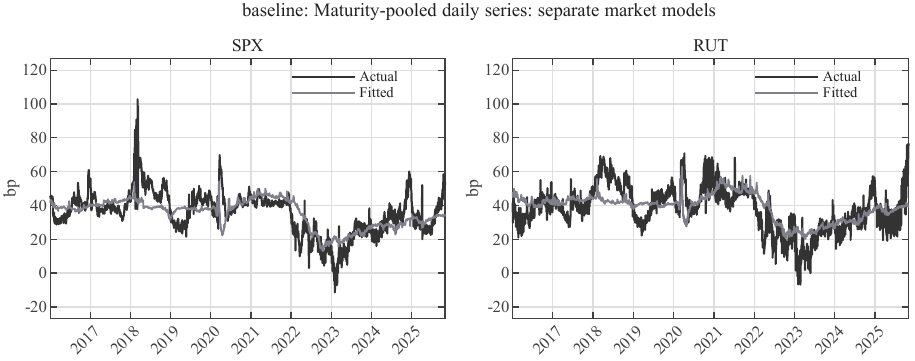}
		\includegraphics[width=6.5in]{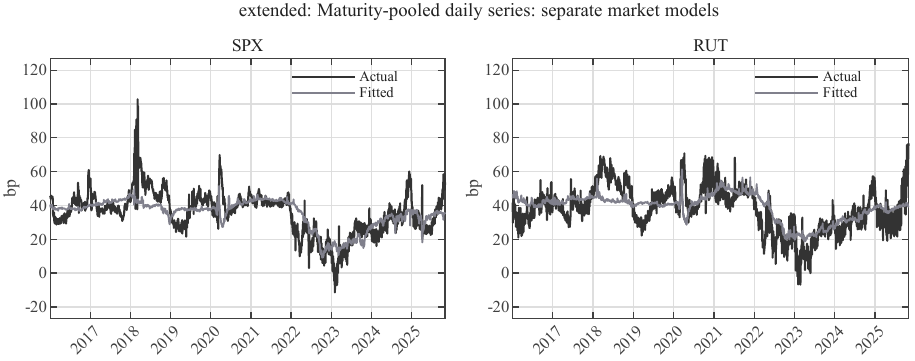}
		\caption{Maturity-pooled daily fit under separate-market regressions:
			baseline versus drift-extended specification. The upper panels report baseline
			fitted values. The lower panels report fitted values from the drift-extended
			specification with \(n=504\) trading days.}
		\label{fig:cg_is}
	\end{figure}
	\FloatBarrier
	
	Figure~\ref{fig:cg_is} compares actual and fitted maturity-pooled daily carry
	gaps. Both specifications capture a substantial part of the low-frequency level
	variation. Adding the drift term improves the alignment between fitted and
	actual carry gaps, especially in SPX. The RUT improvement is in the same
	direction but visibly weaker.
	
	\FloatBarrier
	\begin{table}[H]
		\centering
		\onehalfspacing
		\footnotesize
		\caption{In-sample performance: baseline versus drift-extended specification}
		\label{tab:is_drift_ext}
		\begin{tabular}{lrrrrrrr}
			\toprule
			Market
			& \(R^2\) baseline
			& \(R^2\) ext.
			& \(\Delta R^2\)
			& Adj. \(R^2\) baseline
			& Adj. \(R^2\) ext.
			& \(\Delta\)RMSE (bp)
			& \(\Delta\)MAE (bp) \\
			\midrule
			SPX
			& 0.3124 & 0.3475 & 0.0350
			& 0.3123 & 0.3473
			& -0.341 & -0.364 \\
			RUT
			& 0.2809 & 0.2869 & 0.0060
			& 0.2807 & 0.2867
			& -0.058 & -0.061 \\
			\bottomrule
		\end{tabular}
	\end{table}
	\FloatBarrier
	
	Table~\ref{tab:is_drift_ext} confirms the visual pattern. In SPX, \(R^2\) rises
	from 0.3124 to 0.3475, adjusted \(R^2\) rises from 0.3123 to 0.3473, and both
	RMSE and MAE decline. In RUT, the signs of the changes are the same, but the
	incremental fit is much smaller. The in-sample evidence therefore points to a
	common drift-proxy channel with substantially stronger pass-through in SPX.
	
	To examine the maturity scaling of the \(r\hat{\mu}\tau\) term, I also evaluate
	fit within maturity buckets. This exercise uses the same full-sample
	market-specific coefficients and reports within-bucket fit of the corresponding
	fitted values.
	
	\FloatBarrier
	\begin{table}[H]
		\centering
		\small
		\onehalfspacing
		\caption{In-sample fit by maturity bucket: baseline versus drift-extended specification}
		\label{tab:is_tau_bin}
		\begin{tabular}{llrrrrr}
			\toprule
			Market & Maturity bucket & Obs. & \(R^2\) baseline & \(R^2\) extended & \(\Delta R^2\) & \(\Delta\)RMSE (bp) \\
			\midrule
			SPX & 1--2m   & 2,472 & 0.096 & 0.102 &  0.005 & -0.064 \\
			& 2--3m   & 2,377 & 0.184 & 0.196 &  0.013 & -0.137 \\
			& 3--5m   & 4,501 & 0.283 & 0.310 &  0.027 & -0.255 \\
			& 5--7m   & 3,702 & 0.373 & 0.430 &  0.057 & -0.530 \\
			& 7--10m  & 5,263 & 0.472 & 0.532 &  0.060 & -0.575 \\
			& 10--14m & 6,067 & 0.525 & 0.567 &  0.042 & -0.371 \\
			& 14--21m & 3,751 & 0.306 & 0.354 &  0.049 & -0.462 \\
			& 21m+    & 1,235 & 0.182 & 0.288 &  0.106 & -0.907 \\
			\midrule
			RUT & 1--2m   & 2,471 & 0.112 & 0.113 &  0.001 & -0.010 \\
			& 2--3m   & 2,338 & 0.205 & 0.206 &  0.001 & -0.011 \\
			& 3--5m   & 2,201 & 0.254 & 0.259 &  0.005 & -0.046 \\
			& 5--7m   & 1,934 & 0.231 & 0.259 &  0.028 & -0.229 \\
			& 7--10m  & 2,895 & 0.284 & 0.324 &  0.041 & -0.341 \\
			& 10--14m & 2,683 & 0.472 & 0.493 &  0.020 & -0.206 \\
			& 14--21m & 2,888 & 0.482 & 0.469 & -0.013 &  0.122 \\
			& 21m+    & 1,235 & 0.412 & 0.360 & -0.052 &  0.467 \\
			\bottomrule
		\end{tabular}
		\begin{minipage}{6.5in}
			\footnotesize
			\emph{Notes:} The table reports within-bucket fit using fitted values from the full-sample
			market-specific regressions. The drift-extended specification adds only
			\(GBM^{\hat{\mu},OIS,1Y}_{i,t}\) to the baseline.
			\(\Delta R^2\) and \(\Delta\)RMSE are computed as extended minus baseline.
		\end{minipage}
	\end{table}
	\FloatBarrier
	
	Table~\ref{tab:is_tau_bin} shows that the SPX gains are broad-based and become
	more visible at longer maturities. The \(R^2\) gain rises from 0.005 in the
	1--2 month bucket to 0.106 in the 21m+ bucket, and RMSE falls in every bucket.
	This pattern is consistent with the \(r\hat{\mu}\tau\) structure: if the
	drift-sensitive implementation burden scales with maturity, its empirical
	visibility should increase at longer tenors.
	
	The RUT pattern is weaker and less monotone. The drift-extended specification
	improves fit at short and intermediate maturities, with the largest gain in the
	7--10 month bucket, but underperforms the baseline in the two longest buckets.
	Thus, the maturity-bin evidence reinforces the market-level result. The
	\(\hat{\mu}\)-based channel is strongly visible in SPX, while in RUT it appears
	weaker and concentrated at intermediate maturities.
	
	\subsection{Out-of-Sample Results}
	\label{subsec:oos_results}
	
	I next evaluate the generalizability of the drift-extended specification using
	leave-one-year-out validation. Each calendar year is held out in turn, the
	coefficients are estimated on the remaining years, and fit is evaluated on the
	holdout year.
	
	This design is chosen because the sample is short for a conventional
	expanding-window forecasting exercise. The common OIS-matched sample runs from
	January 2016 to October 2025, leaving fewer than ten full years of data. An
	expanding-window design would therefore depend heavily on a short initial
	training period and a small number of post-training regimes. LOYO validation
	uses the limited sample more symmetrically: every calendar year, including
	stress, transition, and stabilization regimes, is tested as an excluded year.
	
	The LOYO exercise should not be interpreted as a fully causal real-time
	forecasting test. For a given holdout year, the training sample includes
	observations both before and after that year. Its purpose is narrower: to check
	whether the drift-extended relation is driven by overfitting to particular
	calendar regimes, or whether the same coefficient structure remains informative
	when any one year is removed from estimation.
	
	\FloatBarrier
	\begin{figure}[H]
		\centering
		\includegraphics[width=6.5in]{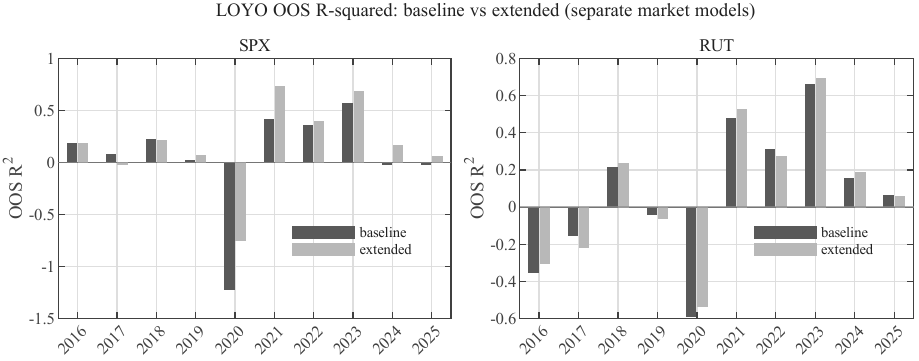}
		\caption{LOYO out-of-sample \(R^2\): baseline versus drift-extended specification under separate-market regressions.
			Bars report holdout-year OOS \(R^2\) for each market.}
		\label{fig:oos}
	\end{figure}
	\FloatBarrier
	
	Figure~\ref{fig:oos} reports year-by-year LOYO out-of-sample \(R^2\). The
	improvement is clear in SPX, especially after 2020. In the 2020 holdout, where
	the baseline fails sharply, OOS \(R^2\) improves from \(-1.221\) to \(-0.748\).
	In 2021, it rises from 0.416 to 0.733. The 2023 and 2024 holdouts also show
	sizable gains. RUT improves overall as well, but the gains are smaller and less
	uniform across years.
	
	\FloatBarrier
	\begin{table}[H]
		\centering
		\small
		\onehalfspacing
		\caption{Year-by-year LOYO out-of-sample \(R^2\): baseline versus drift-extended specification}
		\label{tab:oos_yearly_drift_ext}
		\begin{tabular}{lrrr|rrr}
			\toprule
			& \multicolumn{3}{c|}{SPX} & \multicolumn{3}{c}{RUT} \\
			\cmidrule(lr){2-4} \cmidrule(lr){5-7}
			Holdout year
			& Baseline & Extended & \(\Delta R^2\)
			& Baseline & Extended & \(\Delta R^2\) \\
			\midrule
			2016 &  0.185 &  0.178 & -0.008 & -0.350 & -0.304 &  0.045 \\
			2017 &  0.074 & -0.016 & -0.090 & -0.152 & -0.218 & -0.066 \\
			2018 &  0.221 &  0.211 & -0.010 &  0.213 &  0.233 &  0.020 \\
			2019 &  0.023 &  0.065 &  0.042 & -0.041 & -0.060 & -0.019 \\
			2020 & -1.221 & -0.748 &  0.474 & -0.587 & -0.533 &  0.055 \\
			2021 &  0.416 &  0.733 &  0.317 &  0.477 &  0.522 &  0.045 \\
			2022 &  0.357 &  0.391 &  0.034 &  0.308 &  0.271 & -0.037 \\
			2023 &  0.571 &  0.680 &  0.109 &  0.661 &  0.692 &  0.031 \\
			2024 & -0.016 &  0.161 &  0.177 &  0.153 &  0.187 &  0.034 \\
			2025 & -0.022 &  0.053 &  0.076 &  0.064 &  0.059 & -0.004 \\
			\bottomrule
		\end{tabular}
		\begin{minipage}{6.35in}
			\footnotesize
			\emph{Notes:} The table reports holdout-year OOS \(R^2\) from separate-market LOYO regressions.
			The drift-extended specification adds only \(GBM^{\hat{\mu},OIS,1Y}_{i,t}\) to the baseline.
			\(\Delta R^2\) is computed as extended minus baseline.
		\end{minipage}
	\end{table}
	\FloatBarrier
	
	Table~\ref{tab:oos_yearly_drift_ext} shows that the OOS improvement is not
	driven by a single holdout year. In SPX, the drift-extended specification
	improves OOS \(R^2\) in seven of ten years, with the largest gains in 2020,
	2021, 2023, and 2024. It does not improve every year, especially around
	2016--2018, which confirms that the \(\hat{\mu}\)-based term is not a mechanical
	year-by-year forecasting device. Rather, it improves level alignment in several
	important regimes. In RUT, the extended model improves six of ten years, but the
	gains are generally smaller.
	
	\FloatBarrier
	\begin{table}[H]
		\centering
		\onehalfspacing
		\caption{LOYO out-of-sample performance: baseline versus drift-extended specification}
		\label{tab:oos_drift_ext}
		\begin{tabular}{llrrrr}
			\toprule
			Market & Specification & Mean \(R^2\) & Median \(R^2\) & Pooled \(R^2\) & Mean RMSE (bp) \\
			\midrule
			SPX & Baseline & 0.059 & 0.130 & 0.221 & 13.947 \\
			SPX & Extended & 0.171 & 0.169 & 0.310 & 13.100 \\
			RUT & Baseline & 0.075 & 0.108 & 0.171 & 15.075 \\
			RUT & Extended & 0.085 & 0.123 & 0.188 & 14.930 \\
			\bottomrule
		\end{tabular}
	\end{table}
	\FloatBarrier
	
	Table~\ref{tab:oos_drift_ext} summarizes the LOYO results. In SPX, pooled OOS
	\(R^2\) rises from 0.221 to 0.310, mean OOS \(R^2\) rises from 0.059 to 0.171,
	and mean RMSE falls from 13.947 bp to 13.100 bp. In RUT, pooled OOS \(R^2\)
	increases from 0.171 to 0.188, mean OOS \(R^2\) rises from 0.075 to 0.085, and
	mean RMSE also falls modestly. Thus, the OOS evidence supports the
	drift-extended specification in both markets, but the economic magnitude is much
	more pronounced in SPX.
	
	The gains appear mainly in pooled OOS \(R^2\) and RMSE. This pattern is
	consistent with the role of LOYO as a regime-exclusion diagnostic rather than a
	high-frequency forecasting exercise. The drift-extended specification improves
	level calibration and low-frequency state alignment, consistent with the
	interpretation of \(r\hat{\mu}\tau\) as a reduced-form implementation-risk
	component.
	
	\subsection{Coefficient Structure}
	\label{subsec:coef_results}
	
	I finally examine the coefficient structure of the drift-extended specification.
	As in the baseline analysis, inference is based on date-based HAC
	(Newey--West) standard errors with a maximum lag of 21 trading days.
	
	\FloatBarrier
	\begin{table}[H]
		\centering
		\footnotesize
		\onehalfspacing
		\caption{Regression coefficients: baseline versus drift-extended specification, HAC(21) inference}
		\label{tab:coef_drift_ext}
		\begin{tabular}{lrrrr}
			\toprule
			Regressor & SPX baseline & SPX extended & RUT baseline & RUT extended \\
			\midrule
			Intercept
			& 23.134*** & 27.475*** & 24.577*** & 25.859*** \\
			& (5.709) & (5.454) & (5.403) & (5.459) \\
			
			\(GBM^{\sigma,OIS,1Y}\)
			& -0.548*** & -0.473*** & -0.555*** & -0.467*** \\
			& (0.170) & (0.143) & (0.123) & (0.120) \\
			
			\(GBM^{\sigma,OIS,10Y}\)
			& 0.411** & 0.125 & 0.541*** & 0.406*** \\
			& (0.172) & (0.146) & (0.130) & (0.134) \\
			
			\(GBM^{\hat{\mu},OIS,1Y}\)
			& --- & 0.119*** & --- & 0.054*** \\
			& --- & (0.020) & --- & (0.018) \\
			
			\(BA^{med}/\tau\)
			& 0.256*** & 0.261*** & 0.130*** & 0.135*** \\
			& (0.063) & (0.064) & (0.022) & (0.024) \\
			
			\(NFCI\)
			& -25.839** & -19.577** & -23.961** & -22.286** \\
			& (10.351) & (9.630) & (10.006) & (9.906) \\
			
			\midrule
			\(R^2\)
			& 0.312 & 0.347 & 0.281 & 0.287 \\
			RMSE (bp)
			& 13.199 & 12.858 & 13.951 & 13.893 \\
			MAE (bp)
			& 8.682 & 8.318 & 10.103 & 10.043 \\
			\bottomrule
		\end{tabular}
		\begin{flushleft}
			\footnotesize
			Notes: Standard errors in parentheses are date-based HAC (Newey--West)
			standard errors with maximum lag 21 trading days.
			***, **, and * denote significance at the 1\%, 5\%, and 10\% levels, respectively.
		\end{flushleft}
	\end{table}
	\FloatBarrier
	
	Table~\ref{tab:coef_drift_ext} reports the main coefficient estimates. The key
	result is that \(GBM^{\hat{\mu},OIS,1Y}\) is positive in both markets and
	statistically significant at the 1\% level under HAC(21) inference. The
	coefficient is 0.119 in SPX and 0.054 in RUT. Thus, the \(\hat{\mu}\)-based
	\(r\hat{\mu}\tau\) proxy remains linked to the carry gap after controlling for
	diffusion path risk, trading frictions, and broad financial conditions.
	
	The coefficient is a reduced-form loading on
	\(GBM^{\hat{\mu},OIS,1Y}\), so its economic magnitude depends on how changes in
	\(\hat{\mu}\) map into the regressor:
	\[
	GBM^{\hat{\mu},OIS,1Y}_{i,t}
	=
	10^4
	\left(\frac{OIS^{1Y}_t}{100}\right)
	\hat{\mu}_{i,t}
	\tau_{i,t}.
	\]
	The implied sensitivity of the carry gap to a one-percentage-point increase in
	the drift proxy is therefore
	\[
	\frac{\partial CG^{bp}_{i,t}}{\partial \hat{\mu}^{pp}_{i,t}}
	=
	\hat{\psi}_i
	\cdot
	OIS^{1Y}_{t,\%}
	\cdot
	\tau_{i,t},
	\]
	where \(OIS^{1Y}_{t,\%}\) is the one-year OIS rate in percentage points and
	\(\hat{\mu}^{pp}_{i,t}\) is the drift proxy in percentage points. At
	\(OIS^{1Y}=4\%\) and \(\tau=0.5\), a one-percentage-point increase in
	\(\hat{\mu}\) implies
	\[
	0.119 \times 4 \times 0.5 = 0.24 \text{ bp}
	\]
	for SPX and
	\[
	0.054 \times 4 \times 0.5 = 0.11 \text{ bp}
	\]
	for RUT. A ten-percentage-point increase in the 504-day \(\hat{\mu}\) proxy
	therefore corresponds to roughly 2.4 bp in SPX and 1.1 bp in RUT under the same
	rate and maturity assumptions. This is a low-frequency sensitivity to the
	rolling \(\hat{\mu}\) proxy, not a response to same-day index returns.
	
	The positive sign is consistent with the net-enforcement interpretation. In the
	theoretical argument, a positive true drift \(\mu\) makes the enforcement
	direction containing a short futures leg face a more adverse margin path, while
	the opposite direction benefits from the same drift environment. The observed
	carry gap aggregates these two enforcement directions. The empirical coefficient
	on \(GBM^{\hat{\mu},OIS,1Y}\) should therefore be read as a reduced-form
	pass-through loading that combines directional arbitrage pressure, margin
	burden, and arbitrage-capital supply, not as a structural estimate of \(\mu\).
	
	The behavior of the existing GBM block is also informative. In SPX, adding
	\(GBM^{\hat{\mu},OIS,1Y}\) reduces the coefficient on
	\(GBM^{\sigma,OIS,10Y}\) from 0.411 to 0.125, and the coefficient is no longer
	statistically significant. This suggests that the \(\hat{\mu}\)-based proxy
	absorbs part of the low-frequency variation previously captured by the
	long-horizon OIS path-risk component. In RUT, the coefficient falls from 0.541 to
	0.406 but remains statistically significant, indicating that the \(\hat{\mu}\)
	proxy and the long-rate path-risk channel contain more distinct information. The
	short-horizon OIS term remains negative and significant in both markets, so the
	drift extension does not eliminate the short-rate path-risk channel.
	
	The trading-friction term, \(BA^{med}/\tau\), remains positive and significant
	with little change in magnitude, and \(NFCI\) remains negative and significant
	in both markets. Thus, the added \(\hat{\mu}\)-based term does not simply proxy
	for execution frictions or broad financial conditions. It adds a separate
	low-frequency implementation-risk component.
	
	\FloatBarrier
	\begin{table}[H]
		\centering
		\footnotesize
		\onehalfspacing
		\caption{LOYO coefficient-sign stability for the drift-extended specification}
		\label{tab:coef_stability_drift_ext}
		\begin{tabular}{lcc}
			\toprule
			Regressor & SPX sign stability & RUT sign stability \\
			\midrule
			\(GBM^{\sigma,OIS,1Y}\) & \(-\) in 10/10 & \(-\) in 10/10 \\
			\(GBM^{\sigma,OIS,10Y}\) & \(+\) in 9/10 & \(+\) in 10/10 \\
			\(GBM^{\hat{\mu},OIS,1Y}\) & \(+\) in 10/10 & \(+\) in 10/10 \\
			\(BA^{med}/\tau\) & \(+\) in 10/10 & \(+\) in 10/10 \\
			\(NFCI\) & \(-\) in 10/10 & \(-\) in 10/10 \\
			\bottomrule
		\end{tabular}
	\end{table}
	\FloatBarrier
	
	Table~\ref{tab:coef_stability_drift_ext} reports coefficient-sign stability
	across LOYO training samples. \(GBM^{\hat{\mu},OIS,1Y}\) is positive in all ten
	folds for both SPX and RUT. In SPX, it is significant at the 1\% level in all
	ten folds; in RUT, it is significant in most folds. The other signs are also
	stable: \(GBM^{\sigma,OIS,1Y}\) is always negative, \(BA^{med}/\tau\) is always
	positive, and \(NFCI\) is always negative. \(GBM^{\sigma,OIS,10Y}\) is positive
	in all RUT folds and in 9 of 10 SPX folds, consistent with the weaker role of
	the long-horizon OIS term in the SPX extended specification.
	
	Overall, the coefficient evidence supports the main empirical prediction of the
	drift-preserving extension. The \(r\hat{\mu}\tau\) proxy has a positive and
	stable loading in both markets, but the fit improvement is asymmetric: the
	drift-extended specification substantially improves SPX in both in-sample and
	out-of-sample tests, while in RUT the coefficient remains positive and
	significant but the fit gain is limited. The evidence therefore supports a
	cautious interpretation. Physical drift \(\mu\) does not enter put--call parity
	as a violation of risk-neutral pricing. Instead, the observable \(\hat{\mu}\)
	proxy enters the reduced-form cost of enforcing the parity relation through
	directional margin burden and capital support.
	
	\section{Discussion}
	\label{sec:discussion}
	
	The results do not weaken put--call parity as a terminal-payoff identity.
	They identify a separate implementation layer. Put--call parity remains a
	no-arbitrage relation at maturity, but the position that enforces it must be
	carried along the pre-maturity price path. That path involves variation margin,
	interim gains and losses, funding needs, and finite arbitrage capital.
	
	The carry gap is the carry-space location where this implementation layer can
	remain visible after quoted price-space parity residuals have been compressed.
	The empirical \(r\hat{\mu}\tau\) term is not a claim that physical drift prices
	option payoffs. It is a reduced-form proxy for the directional margin burden
	predicted by the drift-preserving support-capital argument.
	
	\subsection{Parity Enforcement as an Implementation Problem}
	\label{subsec:discussion_noarb}
	\label{subsec:discussion_bsm}
	\label{subsec:discussion_pq}
	
	Standard no-arbitrage logic begins from the condition that two portfolios with
	the same terminal payoff must have the same price. Under frictionless
	assumptions, this logic is complete. Trading is continuous, funding is
	unconstrained, margin constraints are absent, and arbitrage enforcement is
	immediate.
	
	Actual parity enforcement is not frictionless in this sense. An arbitrageur
	must not only reach the terminal payoff. The arbitrageur must also survive the
	path leading to that payoff. Interim losses require variation-margin payments,
	and limited capital can force position reduction before the terminal payoff is
	realized. Identical terminal payoffs therefore need not imply identical
	implementation paths.
	
	This distinction is especially relevant for the two put--call parity
	enforcement directions,
	\[
	+C-P-F
	\quad\text{and}\quad
	-C+P+F.
	\]
	They enforce the same terminal relation from opposite sides, but their
	futures-leg exposures have opposite signs. In the theoretical argument, when
	\(\mu>0\), the direction containing a short futures leg faces a more adverse
	margin path. When \(\mu<0\), the burden can shift to the opposite direction.
	This asymmetry disappears at the terminal payoff level, but it remains relevant
	over the path on which the position must be financed and maintained.
	
	The baseline diffusion term,
	\[
	r\sigma\sqrt{\tau},
	\]
	summarizes the support-capital burden generated by random path variation. The
	drift-preserving extension adds the empirical term
	\[
	r\hat{\mu}\tau,
	\]
	which proxies for directional margin burden. The positive and stable loading on
	\(GBM^{\hat{\mu},OIS,1Y}\) suggests that this empirical proxy contains
	incremental information about the cost of enforcing the parity relation after
	controlling for diffusion path risk, trading frictions, and financial
	conditions.
	
	In this sense, the ``\(P\) behind \(Q\)'' is not a claim that risk-neutral
	pricing collapses. Put--call parity is a \(Q\)-measure terminal-payoff relation.
	The claim is narrower: the capital-support technology required to implement that
	relation is realized along the physical price path. The observed carry-gap
	dynamics therefore appear partly aligned with an empirical \(P\)-measure drift
	proxy, even though the terminal payoff identity itself remains intact.
	
	\subsection{Joint Enforcement Across Parity Relations}
	\label{subsec:discussion_joint_parity}
	
	The implementation-cost interpretation also explains why quoted put--call
	parity can appear tightly closed while a related forward-channel wedge remains
	visible. In a frictionless benchmark, put--call parity, synthetic-forward
	identification, the traded-futures channel, and the benchmark discounting curve
	are jointly pinned down by the same no-arbitrage logic. If trading is costless,
	funding is unconstrained, and arbitrage capital is unlimited, the call--put
	spread, the traded forward channel, and the benchmark discount curve should all
	be mutually aligned.
	
	With implementation frictions, however, these equalities are not jointly
	costless to enforce. The risk associated with closing one relation must be borne
	by some balance sheet, margin account, funding line, or intermediary capital
	constraint. Arbitrage pressure may therefore close the relation that is cheapest
	or most visible to enforce while leaving a wedge in a related channel that is
	more capital-intensive or institutionally costly.
	
	This provides a natural interpretation of the empirical pattern motivating this
	paper. Quoted put--call parity is the more direct option-market relation and can
	be compressed within a relatively tight option-market enforcement channel. By
	contrast, aligning the OIS-discounted synthetic-forward channel with the
	traded-futures channel can require broader balance-sheet capacity, including
	futures exposure, margin financing, dividend and settlement risks, inventory
	constraints, and funding liquidity. Finite arbitrage capital may therefore be
	allocated first to closing the visible put--call parity residual, while the
	remaining implementation burden appears as a wedge in the synthetic--traded
	forward or discounting channel.
	
	Under this interpretation, the carry gap is not evidence that put--call parity
	itself fails. Rather, the option-implied discount factor need not coincide
	exactly with the OIS benchmark discount factor once implementation costs are
	reflected in the enforcement channel. The difference can be interpreted as an
	implementation premium: compensation for funding costs, path risk, trading
	frictions, and capital usage not captured by the frictionless OIS benchmark.
	
	This interpretation should not be read as a literal spot--futures parity
	violation. For index products, contract design, dividends, settlement, and
	timing conventions make the textbook relation difficult to map one-for-one into
	the data. The conservative claim is that the option-implied synthetic-forward
	channel and the traded-futures/OIS benchmark channel are not perfectly aligned
	after implementation costs. The results suggest that part of this wedge is
	related to volatility, funding conditions, trading frictions, and the
	drift-sensitive margin-burden proxy captured by \(r\hat{\mu}\tau\).
	
	\subsection{A Price-Space Hurdle Interpretation}
	\label{subsec:discussion_price_hurdle}
	
	Although the empirical analysis is conducted in carry space, the fitted carry
	gap has a direct price-space interpretation. Let
	\(\widehat{CG}_{i,t}(T)\) denote the model-implied annualized carry gap in
	decimal units for maturity \(T\), and let \(\tau_{i,t}=T-t\). Since
	\[
	CG_{i,t}(T)
	=
	\frac{1}{\tau_{i,t}}
	\log
	\left(
	\frac{D^{OIS}_{t}(T)}{\hat{B}_{t}(T)}
	\right),
	\]
	the fitted unannualized discount-factor log wedge is
	\[
	\widehat{g}_{i,t}(T)
	=
	\tau_{i,t}\widehat{CG}_{i,t}(T).
	\]
	The corresponding proportional discounting wedge is therefore
	\[
	\exp\{\tau_{i,t}\widehat{CG}_{i,t}(T)\}-1.
	\]
	
	Multiplying this wedge by the relevant forward scale gives a price-space
	implementation hurdle:
	\[
	\widehat{H}_{i,t}(T)
	=
	F_{i,t}(T)
	\left[
	\exp\{\tau_{i,t}\widehat{CG}_{i,t}(T)\}-1
	\right].
	\]
	For small carry gaps,
	\[
	\widehat{H}_{i,t}(T)
	\approx
	F_{i,t}(T)\tau_{i,t}\widehat{CG}_{i,t}(T).
	\]
	If the fitted carry gap is expressed in basis points, this becomes
	\[
	\widehat{H}_{i,t}(T)
	\approx
	F_{i,t}(T)\tau_{i,t}
	\frac{\widehat{CG}^{bp}_{i,t}(T)}{10^4}.
	\]
	
	This transformation gives the model a practical interpretation. A put--call
	parity arbitrageur should not compare a quoted price-space residual only with
	zero. The relevant comparison is between the observed residual and the
	state-dependent implementation hurdle implied by funding costs, volatility,
	maturity, trading frictions, financial conditions, and the \(\hat{\mu}\)-based
	directional margin-burden proxy. If the observed residual is smaller than this
	hurdle, the absence of arbitrage can be consistent with rational implementation
	costs. If it substantially exceeds the hurdle, the trade may offer compensation
	beyond the estimated support-capital burden.
	
	This is not a mechanical trading rule. Actual profitability also depends on
	bid--ask spreads, execution risk, margin rules, financing spreads, balance-sheet
	costs, inventory constraints, and model error. The fitted carry gap should be
	read as a reduced-form screening heuristic: a state-dependent price-space
	benchmark for evaluating whether a parity residual is large enough to compensate
	for the estimated cost of enforcement.
	
	\subsection{Limitations of the Interpretation}
	\label{subsec:discussion_limits}
	
	The interpretation has three main limitations. First, the \(r\hat{\mu}\tau\)
	term is a reduced-form empirical regressor, not a structurally identified
	primitive. The true physical drift \(\mu\) is unobserved. The proxy
	\(\hat{\mu}\) is constructed from a prior rolling OLS slope of the log
	total-return path. The coefficient on \(GBM^{\hat{\mu},OIS,1Y}\) is therefore
	not an estimate of the expected return perceived by arbitrageurs, an equilibrium
	risk premium, or a structural drift parameter. It is a reduced-form loading on
	an observable historical-drift proxy.
	
	Second, the explanatory power of the \(\hat{\mu}\)-based term is not identical
	across markets. In SPX, the improvement in both in-sample and leave-one-year-out
	out-of-sample performance is clear. In RUT, the sign and statistical
	significance of the coefficient are stable, but the incremental fit gain is
	smaller. This suggests that the same directional margin-burden mechanism can
	have different pass-through across markets, depending on liquidity, investor
	composition, maturity structure, and the nature of enforcement capital.
	
	Third, the paper does not propose a replacement for the
	Black--Scholes--Merton framework or for standard no-arbitrage pricing. Its
	contribution is narrower. If the standard model provides the frictionless
	benchmark, this paper shows that the path-dependent capital burden of
	implementing the benchmark relation is empirically visible. A full structural
	model of parity-enforcement pricing, with endogenous margin constraints,
	funding constraints, direction-specific capital burdens, and a sharper mapping
	between true drift \(\mu\) and empirical proxies \(\hat{\mu}\), is left for
	future research.
	
	\section{Conclusion}
	\label{sec:conclusion}
	
	This paper studies put--call parity as an enforced no-arbitrage relation, not
	only as a terminal-payoff identity. The terminal identity remains intact. The
	implementation question is different: the position that enforces parity must be
	carried through the pre-maturity path and is exposed to variation margin,
	interim gains and losses, trading frictions, funding conditions, and finite
	arbitrage capital. The carry gap is the carry-space location where this
	implementation burden can remain visible after quoted price-space parity
	residuals have been compressed.
	
	The paper makes three points. First, it reframes the carry gap as an
	implementation wedge rather than as a failure of put--call parity. In a
	frictionless benchmark, put--call parity, the synthetic-forward channel, the
	traded-futures channel, and the benchmark discounting curve are jointly aligned.
	In actual markets, these relations are not jointly costless to enforce. The
	option-implied discount factor can therefore differ from the OIS benchmark
	discount factor because the full enforcement channel uses margin, funding, and
	balance-sheet capacity.
	
	Second, the paper derives and tests a drift-preserving extension of the GBM
	implementation-risk term. Under zero-drift Brownian motion, the support-capital
	burden scales with \(r\sigma\sqrt{\tau}\). When the true physical drift \(\mu\)
	is preserved in the approximation, a first-order directional margin-burden
	component proportional to \(r\mu\tau\) appears. Since \(\mu\) is unobserved, the
	empirical specification uses \(\hat{\mu}\), a prior rolling-OLS slope of past log
	total-return paths, and implements the drift component as
	\(GBM^{\hat{\mu},OIS,1Y}\). This term is a reduced-form historical-drift proxy,
	not an observed expected return or a structural estimate of \(\mu\).
	
	Third, the evidence shows that the \(\hat{\mu}\)-based term has incremental
	explanatory power for the carry gap after controlling for diffusion path risk,
	trading frictions, and broad financial conditions. The result is strongest in
	SPX, where the extended specification improves both in-sample fit and
	leave-one-year-out validation, and where the maturity-bin pattern is consistent
	with the \(\tau\)-scaled structure of \(r\hat{\mu}\tau\). In RUT, the coefficient
	on \(GBM^{\hat{\mu},OIS,1Y}\) remains positive and stable, but the incremental
	fit gain is smaller and less uniform. The evidence therefore supports a common
	directional margin-burden channel with market-specific pass-through.
	
	These findings do not imply that physical drift prices option payoffs directly,
	nor do they challenge risk-neutral pricing. Put--call parity remains a
	risk-neutral terminal-payoff relation. The result is about the capital-using
	process that enforces that relation before maturity. A parity trade may have a
	deterministic terminal payoff, but the capital required to keep the trade alive
	is committed along the physical price path. In this sense, the ``\(P\) behind
	\(Q\)'' is the implementation layer behind a risk-neutral parity relation.
	
	The fitted carry gap can also be read as a state-dependent price-space hurdle:
	a reduced-form benchmark for whether a quoted parity residual is large enough to
	compensate for estimated enforcement costs. This is not a mechanical trading
	rule. It is an implication of the same implementation-risk interpretation.
	
	A full structural model of parity-enforcement pricing remains for future work.
	Such a model would endogenize margin constraints, funding constraints,
	direction-specific capital burdens, and the allocation of finite arbitrage
	capital across related parity relations. It would also need a sharper mapping
	between the true drift \(\mu\) and empirical proxies \(\hat{\mu}\). The evidence
	in this paper is a reduced-form first step: the terminal parity identity belongs
	to the risk-neutral pricing relation, while the cost of enforcing it is partly
	aligned with physical-path information.
	
	\subsection*{Funding}
	This research did not receive any specific grant from funding agencies in the public, commercial, or not-for-profit sectors.
	
	\subsection*{Declaration of AI usage in manuscript preparation}
	During the preparation of this manuscript, the author used ChatGPT (OpenAI) and Claude (Anthropic) for language refinement and structural clarity.
	All outputs were reviewed and edited by the author, who takes full responsibility for the content.
	
	\subsection*{Declaration of interest}
	The author declares no competing interests.
	
	\newpage
	\onehalfspacing

	\begin{appendices}
		\onehalfspacing
		
		\newpage
		\section{Implementation Details of the Pipeline}
		\label{app:ab21_pipeline}
		
		This appendix summarizes the implementation procedure used to identify option-implied discount factors and construct the carry gap.
		The main identification logic and regression specifications are described in the text.
		Here I focus on data processing and sample construction.
		
		\subsection{Data and Sample}
		
		I identify option-implied discount factors from SPX and RUT European-style index options using the synthetic-forward procedure of \citet{AB21}.
		The option data are minute-level NBBO quotes from ThetaData.
		The sample is restricted to the common period in which both option quotes and the OIS curve are observed.
		Because SPX and RUT options are European-style index options, early-exercise premia do not directly enter the put--call-parity-based identification.
		
		\subsection{Identification of Option-Implied Discount Factors}
		
		For each date--maturity cell, I form call--put pairs with the same strike and maturity.
		I then compute the synthetic forward as the difference between the call and put mid quotes.
		The option-implied discount factor and forward value are jointly recovered by estimating the linear relation between synthetic forwards and strikes across the strike cross-section.
		
		This procedure has two advantages.
		First, it does not rely on a single ATM contract or an arbitrary moneyness interval, but uses the cross-section of options at the same maturity.
		Second, because the forward value is identified jointly from option prices, the procedure does not require a separate dividend estimate as a direct external input.
		It also mitigates nonsynchronicity problems that can arise when spot, futures, dividend, and interest-rate data are combined from separate sources.
		
		\subsection{OIS Benchmark and Carry Gap}
		
		The benchmark discount factor is constructed from the daily OIS curve.
		I apply a standard bootstrapping procedure to recover discount factors by maturity and then match the OIS discount factor to each option maturity.
		As defined in the main text, the carry gap is the annualized log wedge between the option-implied discount factor and the OIS discount factor.
		The empirical analysis uses the basis-point-scaled version of this wedge.
		
		\subsection{Sample Filters and Daily Aggregation}
		
		The preprocessing focuses on minimal filters that improve the stability of cross-sectional identification.
		I exclude observations without a valid call--put pair at the same strike and maturity.
		I also remove quotes with extremely low prices or excessive bid--ask spreads.
		Date--maturity cells with too few valid strikes are excluded, as are dates on which OIS curve construction fails.
		
		The final sample is a market--date--maturity panel in which both the option-implied discount factor and the matched OIS benchmark discount factor can be constructed reliably.
		For figures and diagnostic exercises that require daily market-level series, I aggregate valid maturity cells within each date by taking the median carry gap.
		Median aggregation reduces the influence of outliers and transitory quote noise.
		
		\subsection{Drift Proxy and Regression Evaluation}
		
		The physical-drift proxy used in the drift-extended specification is constructed using the prior-only rolling OLS slope procedure described in the main text.
		For date \(t\), the drift proxy is computed only from the log total-return path through \(t-1\), which prevents look-ahead bias.
		The main specification applies a common 504-trading-day lookback horizon to both SPX and RUT.
		
		Regression inference uses date-based HAC(Newey--West) standard errors.
		The main results use a maximum lag of 21 trading days.
		Out-of-sample performance is evaluated using leave-one-year-out validation.
		For each calendar year, I exclude that year as the holdout sample, estimate the regression coefficients on the remaining years, and compute out-of-sample \(R^2\), RMSE, MAE, and fitted--actual correlation in the excluded year.
		This procedure tests whether the improvement from the drift-preserving term is only an in-sample fit gain or whether it is repeatedly preserved across year-by-year sample splits.
		
	\end{appendices}
	
	\end{document}